\documentclass[a4paper]{article}

\usepackage{indentfirst}
\usepackage{fullpage}
\usepackage{amsmath}
\usepackage{amsthm}
\usepackage{latexsym}
\usepackage{epsf}
\usepackage{graphicx, color}
\usepackage[lofdepth,lotdepth]{subfig}
\usepackage{subfloat}

\makeatletter
\renewcommand\paragraph{%
  \@startsection{paragraph}{4}{0mm}%
   {-\baselineskip}%
   {.5\baselineskip}%
   {\normalfont\normalsize\bfseries}}
\makeatother

\date{}

\begin{document}

\title{\Large \bf \boldmath\ The dynamics of inextensible capsules \\in shear flow  under the effect of the nature state}

\author{
Xiting Niu, Lingling Shi, Tsorng-Whay Pan\footnotemark[1], Roland Glowinski\\
Department of Mathematics, University of Houston, Houston, TX 77204, USA}

\renewcommand{\thefootnote}{\fnsymbol{footnote}}
\footnotetext[1]{Corresponding author: pan@math.uh.edu}
\newtheorem{remark}{Remark}[section]

\newcommand{\bff}{{\bf f}}
\newcommand{\bn}{{\bf n}}
\newcommand{\bnabla}{{\boldsymbol{\nabla}}}
\newcommand{\bg}{{\bf g}}
\newcommand{\bu}{{\bf u}}
\newcommand{\bD}{{\bf D}}
\newcommand{\bF}{{\bf F}}
\newcommand{\bG}{{\bf G}}
\newcommand{\bU}{{\bf U}}
\newcommand{\bv}{{\bf v}}
\newcommand{\bx}{{\bf x}}
\newcommand{\bV}{{\bf V}}
\newcommand{\bz}{{\bf 0}}
\newcommand{\cth}{{\mathcal{T}_h}}
\newcommand{\calt}{{\mathcal{T}}}
\newcommand{\g}{{\bf g}}
\newcommand{\gx}{{\Gamma_{-} (0, \Delta t)}}
\newcommand{\Gx}{{{\overrightarrow{Gx}}^{\perp}}}
\newcommand{\dsum}{{\displaystyle\sum}}
\newcommand{\into}{{\displaystyle{\int_{\Omega}}}}
\newcommand{\intG}{{\displaystyle{\int_{\Gamma}}}}
\newcommand{\oo}{{\overline{\Omega}}}
\newcommand{\ox}{{\Omega \times (0, \Delta t)}}
\newcommand{\oxot}{{\Omega \times (0,T)}}
\newcommand{\obo}{{\Omega \backslash \overline{B(0)}}}
\newcommand{\obt}{{\Omega \backslash \overline{B(t)}}}
\newcommand{\R}{{I\!\!R}}
\newcommand{\blambda}{{\boldsymbol{\lambda}}}
\newcommand{\bmu}{{\boldsymbol{\mu}}}
\newcommand{\bsigma}{{\boldsymbol{\sigma}}}

\maketitle

\begin{abstract}

The effect of the nature state on the motion of an inextensible capsule in simple shear flow  has been studied 
in this paper. Besides the viscosity ratio of the internal fluid and external fluid of the capsule, 
the nature state effect also plays a role for having the transition between two well known motions, tumbling  and 
tank-treading (TT) with the long axis oscillating about a fixed inclination angle (a swinging mode), when varying 
the shear rate. The intermittent region between tumbling and TT with a swinging mode of the capsule with 
a biconcave rest shape has been obtained in a narrow range of the capillary number. In such region, the 
dynamics of the capsule is a mixture of tumbling and TT with a swinging mode; when having the tumbling motion, 
the membrane tank-tread backward and forward within a small range. As the capillary number 
is very close to and below the threshold for the pure TT with a swinging mode, the capsule tumbles once after 
several TT periods in each cycle. The number of TT periods in one cycle decreases 
with respect to the decreasing of the capillary number, until the capsule has one tumble  and one TT period 
alternatively and such alternating motion exists over a range of the capillary number; and then the capsule
performs more tumbling between two consecutive TT periods when reducing the capillary number further, and finally shows 
pure tumbling. The critical value of the swelling ratio for having the intermittent region has been estimated. 

\end{abstract}

\baselineskip 14pt

\setlength{\parindent}{1.5em}

\setcounter{section}{0}

\vskip 2ex

\section{Introduction}
The dynamical behavior of deformable entities such as lipid vesicles, capsules,
and red blood cells (RBCs) in flows has received increasing attention experimentally, 
theoretically, and numerically in recent years. Lipid vesicles (\cite{Hass1997}--\cite{Noguchi2007}),
non-spherical capsules (\cite{Keller1982}--\cite{Dupont2013}),
and red blood cells (\cite{Fischer1978}--\cite{Fischer2004}) 
show phenomenologically similar behavior in shear flows, which are (i) an unsteady tumbling motion,
(ii) a tank--treading rotation with a stationary shape and a fixed inclination angle with respect to the
flow direction, and (iii) while tank-treading (TT), the long axis oscillates about a fixed inclination angle 
which is called swinging mode in \cite{Noguchi2007} (but it is still called the tank--treading in \cite{Skotheim2007}).
The aforementioned motions depend on the shear rate, viscosity ratio between internal fluid and external
fluid, membrane viscosity and other parameters.
Noguchi and Gompper \cite{Noguchi2004, Noguchi2005, Noguchi2007} have studied the dynamics of vesicles
in simple shear flow using mesoscale simulations of dynamically triangulated surfaces. Vesicles are
found to transit from steady tank--treading to unsteady tumbling, or TT with a swinging mode with increasing membrane 
viscosity. In \cite{Noguchi2007}, they have developed a model based on the classical 
model of Keller and Skalak \cite{Keller1982} (KS model) for the membrane to theoretically explain the 
vesicle motion. Using such model, they obtained the swinging mode and studied the dependency 
of transition between tumbling and TT with a swinging mode on the shear rate, the viscosity ratio of the membrane and
the internal fluid, and the reduced volume \cite{Noguchi2007}.  
Abkarian {\it et al.} \cite{Abkarian2007} observed the intermediate motions at the transition from 
swinging to tumbling (resp., tumbling to swinging) by reducing (resp., increasing) the flow shear rate. 
A simplified model with a fixed elliptical shape for cell membrane has been studied to support 
their observations.
In \cite{Skotheim2007}, Skotheim and Secomb introduced an elastic energy term based on the phase angle
of the tank--treading rotation   to the KS model. They observed tumbling, tank--treading 
(with a swinging mode), and the intermittent behavior at the transition between tumbling and tank--treading 
and analyzed the influence of the viscosity ratio, membrane elasticity, shape, and the shear rate on the 
motion of a capsule of either prolate or oblate shape.
The both models considered in \cite{Skotheim2007,Abkarian2007} take into account the membrane elastic energy.
Tsubota {\it et al.} \cite{Tsubota2010} used a spring model fully coupled with fluid flow to study cell motion
dependency on the natural state of membrane through the bending energy term in two--dimensional shear flow.
When being at uniform natural state (i.e., all reference angles are set to be the same), the cell appears to 
tank--tread with an inclination angle unchanged and independent of the preset value of reference angles. 
But when a biconcave resting shape is assumed as the natural state (called non--uniform state), the cell is 
observed to perform tumbling and tank--treading with a swinging mode.  Tsubota
{\it et al.}  \cite{Tsubota2010} believed that the intermittency between tumbling and TT with a swinging mode  
would occur in very narrow range of parameter space; but they did not obtain such intermittent region
because, as discussed later in this paper, the swelling ratios of their cells are not small enough. The elastic energy term 
introduced by Skotheim and Secomb in \cite{Skotheim2007} has a similar characteristics like the bending
energy defined with a non--uniform state used by Tsubota {\it et al.} in \cite{Tsubota2010} since both 
have their preferred rest states, which are preferred phase angles at zero and 180 degrees and a biconcave resting shape
for Skotheim and Secomb's model and Tsubota {\it et al.}'s model, respectively.
In \cite{Young2011}, Vlahovska {\it et al.} used perturbation approach to study the motion of an almost 
spherical capsule in shear flow at the Stokes regime. Their reduced models are more general and have no restriction 
of the fixed shape on the capsule when comparing with the one used in \cite{Skotheim2007}. Without the elastic energy 
introduced by Skotheim and Secomb in \cite{Skotheim2007}, Vlahovska {\it et al.} obtained the intermittent behavior when
having deformation only in the shear plane and that the intermittent behavior disappears in Stokes flow for an almost 
spherical capsule which can deform in the vorticity direction.
They concluded that the intermittency is an artifact of the shape preservation. But they did not analyzed the cases
of capsules of more interesting shapes such as a biconcave shape. 
In \cite{Tsubota2013}, the transition between tumbling and tank--treading motions has been considered
while studying the model of the RBC membrane fully coupled with the Stokes flow. Tsubota  {\it et al.} obtained 
no intermittent dynamics around the transition between tumbling and tank--treading motions for the cases of 
the viscosity ratio, $\lambda$, of internal and external fluids between 0.1 and 0.3, which is quite low 
when comparing with the values used in   \cite{Skotheim2007, Young2011}.
In \cite{Fedosov2010}, Fedosov  {\it et al.} used a two--dimensional spring network with the natural state 
in their bending energy term to model red blood cell membrane. They obtained the intermittent region of mixed 
dynamics in shear flow for the cases of the viscosity ratio $\lambda$ greater than or equal to 1. 

In this paper, we have analyzed the capsule motion in simple shear flow via direct numerical simulations under the 
effect of the natural state. In \cite{Skotheim2007}, Skotheim and Secomb restricted the capsules to stay in the shear plane 
with the fixed shape and obtained the intermittent  behavior by studying the in--plane inclination angle and phase angle 
with the membrane elastic energy, which is actually a two--dimensional model.  Their membrane elastic energy term, which is based on 
the phase angle, has preferred phase angles at zero and 180 degrees; i.e., the point at the rim likes to go back to the rim. 
Our bending elastic energy term based on the natural state also has such preference. In simulations, we have relaxed the restriction 
of the non--deformability (i.e., the fixed shape) and have considered a fully interaction of the fluid and capsule in two dimensional 
shear flow. Furthermore, we have studied the dependency of the capsule motion on the capillary number $C_a=\mu \dot{\gamma} R_0^3/B$,  
where $\mu$, $\dot{\gamma}$, $R_0$, and $B$ stand for the fluid viscosity, the shear rate of fluid flow  based on the 
gradient of the velocity at the wall, the effective radius of the  capsule and the bending modulus, respectively.
We have obtained that the swelling ratio of the capsule does have its effect on the intermittent region.
The intermittent region of mixed dynamics between tumbling and TT with a swinging mode of the capsule with 
a biconcave rest shape has been obtained in a narrow range of the capillary number.  For the capsule 
of swelling ratio greater than 0.6, it is almost impossible to capture the intermittent region computationally since the size of the 
capillary number range for such region is about zero if it exists. Our results are consistent with 
the results obtained by Tsubota {\it et al.}  in \cite{Tsubota2010} since the cells used in their simulations have the 
swelling ratio of 0.7.

The contents of this paper are as follows: We discuss the models and numerical methods briefly in Section 2.
In Section 3, we show validations and also carry out the go--and--stop experiments on the model to illustrate the effect
the natural state  on the preferred rest shape. The dynamics of an inextensible capsule under the effect of the natural state 
in shear flow  are studied at  different values of the capillary number and confinement ratio in Section 4.
In Section 5, the details on the dynamics in the intermittent region are studied. 
The conclusions are summarized in Section 6.

\section{Models and numerical methods}

\begin{figure}[!htp]
\begin{center}
\subfloat[][]{\includegraphics[width=3.34in]{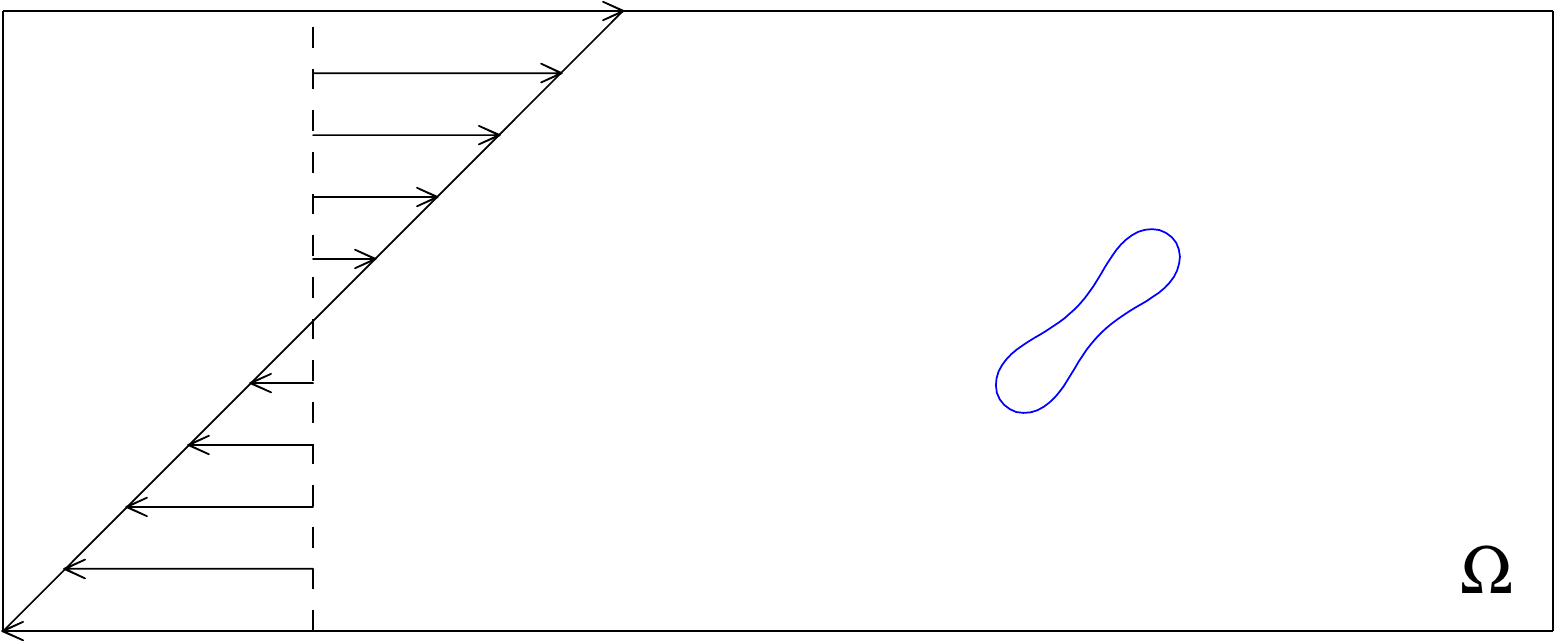}\label{FIG.1.1}}
\subfloat[][]{\includegraphics[width=1.38in]{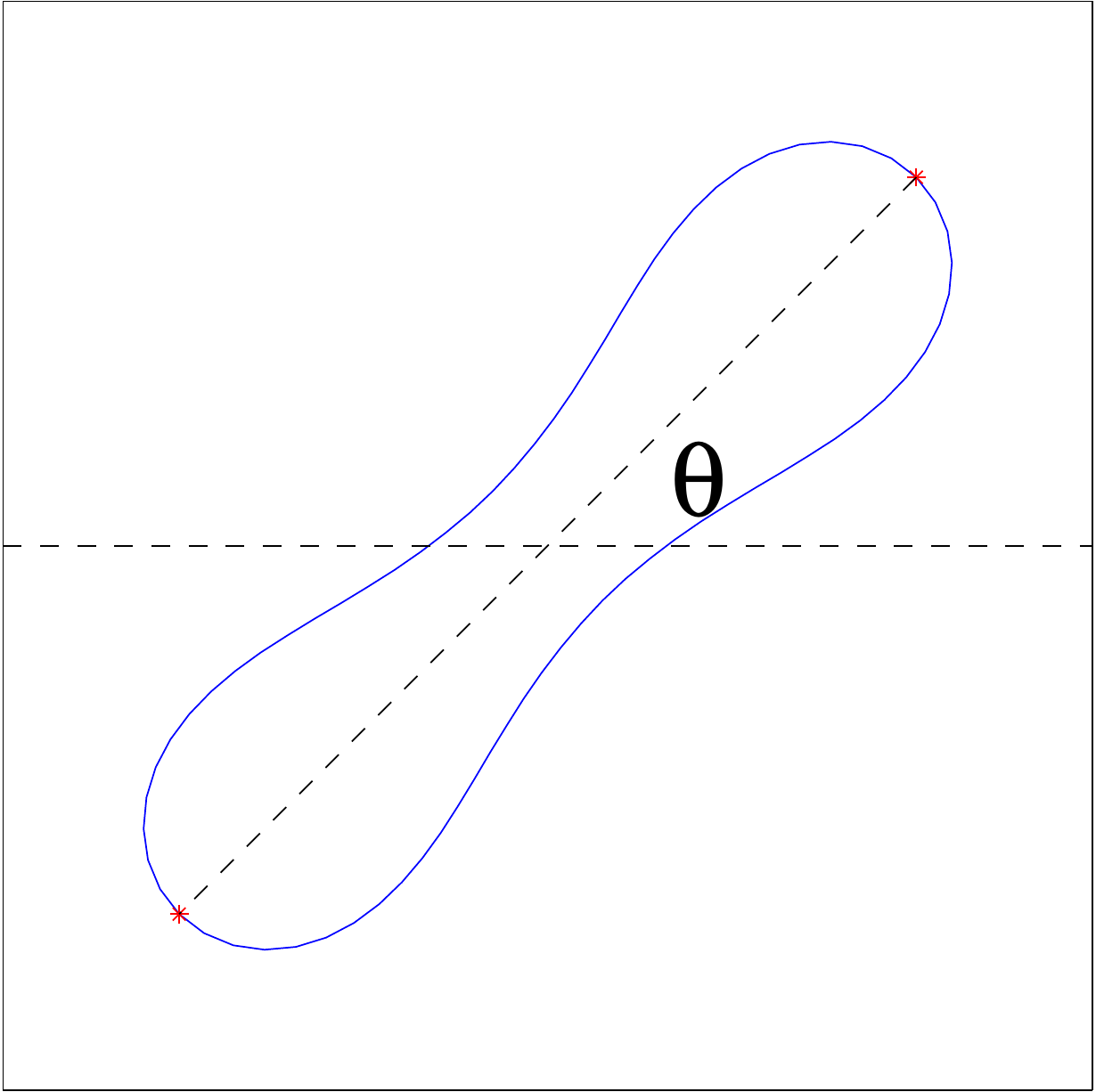}\label{FIG.1.2}}\\
\subfloat[][]{\includegraphics[width=4.72in]{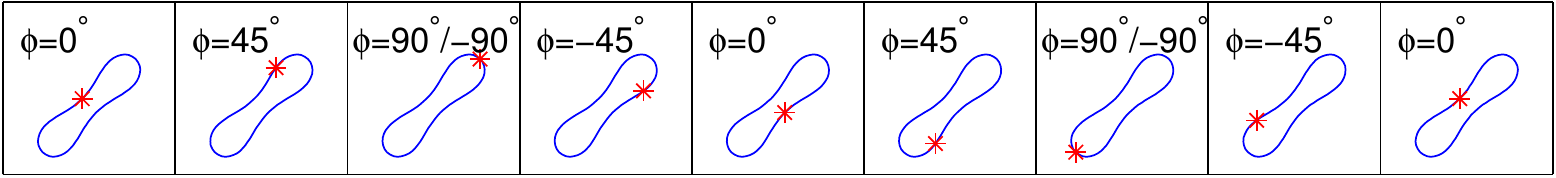}\label{FIG.1.3}}
\end{center}
\caption{(Color online) Schematic diagram of \protect\subref{FIG.1.1} an inextensible biconcave capsule
in shear flow with the computational domain $\Omega$,
\protect\subref{FIG.1.2} the inclination angle $\theta$, 
and \protect\subref{FIG.1.3}
the phase angle $\phi$.} \label{FIG.1}
\end{figure}

A capsule with non-spherical rest shape is suspended in a
domain $\Omega$ filled with a fluid which is incompressible and Newtonian as in Figure \ref{FIG.1}.
The inclination angle $\theta$ and phase angle $\phi$ are defined as in Figure \ref{FIG.1.2} and \ref{FIG.1.3}.
respectively. For some $T>0$, the
governing equations for the fluid--capsule system are the Navier--Stokes equations

\begin{eqnarray}
&&\rho_f \left( \dfrac{\partial \bu}{\partial t} + \bu \cdot\nabla \bu \right) =-\nabla p+\mu \triangle \bu+\bff_B,
\text{    in   }  \Omega \times (0, T),   \label{eqn:1a}  \\
&& \nabla \cdot \bu=0,\text{    in   }  \Omega \times (0, T).  \label{eqn:1b}
\end{eqnarray}

\noindent with the following boundary and initial conditions:
\begin{eqnarray}
&& \bu={\bu_{max}} \text{ on the top and } {-\bu_{max}}\text{ on the bottom of $\Omega$ and $\bu$ 
is periodic in the $x$ direction,} \label{eqn:1c}  \\
&& \bu(\bx,0)=\bu_0(\bx) ,  \text{    in   }  \Omega  \label{eqn:1d}
\end{eqnarray}
\noindent where $\bu$ and $p$ are the fluid velocity and pressure, respectively,
$\rho_f$ is the fluid density, and $\mu$ is the fluid viscosity, which is assumed to be
constant for the entire fluid. In (\ref{eqn:1a}), $\bff_B$ accounts for the force acting on the fluid/capsule interface. 
The boundary condition in (\ref{eqn:1c}) is ${\bu_{max}} =(U,0)^t$ for simple shear flow.
In (\ref{eqn:1d}), $\bu_0(\bx)$ is the initial fluid velocity. 

\begin{figure}
\begin{center}
\includegraphics[width=2in]{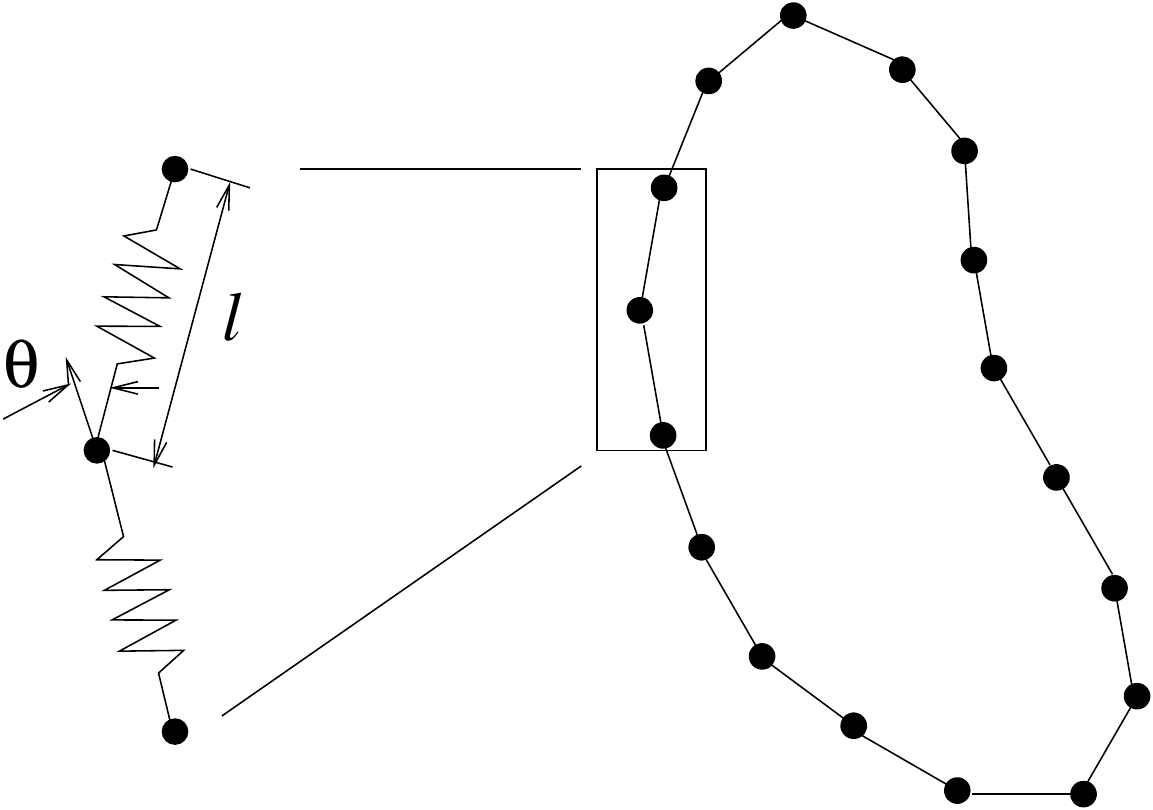}
\end{center}
\caption{The elastic spring model of the capsule membrane.} \label{FIG.2}
\end{figure}

\subsection{Elastic spring models}

An elastic spring model similar to the one  used in \cite{Tsubota2010} is considered in this paper to describe
the deformable behavior and elasticity of capsules. Based on this model, the capsule membrane can be 
viewed as membrane particles connecting with the neighboring membrane particles by springs, as shown in 
Figure \ref{FIG.2}. Energy stores in the spring due to the change of the length $l$ of the spring with 
respect to its reference length $l_0$ and the change in angle $\theta$ between two neighboring springs. 
The total energy per unit thickness of the capsule membrane, $E=E_l+E_b$, is the sum of the one for 
stretch and compression and the one for the bending which, in particular, are

\begin{equation}
E_{l}=\frac{k_{l}}{2}\sum_{i=1}^{N}(\frac{l_{i}-l_{0}}{l_{0}})^{2}, \ \ 
E_{b}=\frac{k_{b}}{2}\sum_{i=1}^{N}\tan^{2}(\frac{\theta_{i}-\theta_{i}^{0}}{2}). \label{eqn:2}
\end{equation}
In equation (\ref{eqn:2}), $N$ is the total number of
the spring elements, and $k_{l}$ and $k_{b}$ are spring constants
for changes in length and bending angle, respectively. The set of reference angles $\{\theta_i^0\}_{i=1}^N$ 
corresponds to a preset natural state, where $\theta_i^0=constant$ for all $i$ corresponds to a uniform 
natural state, and otherwise a nonuniform natural state. 

\begin{figure}
\begin{center}
\includegraphics[width=2.25in]{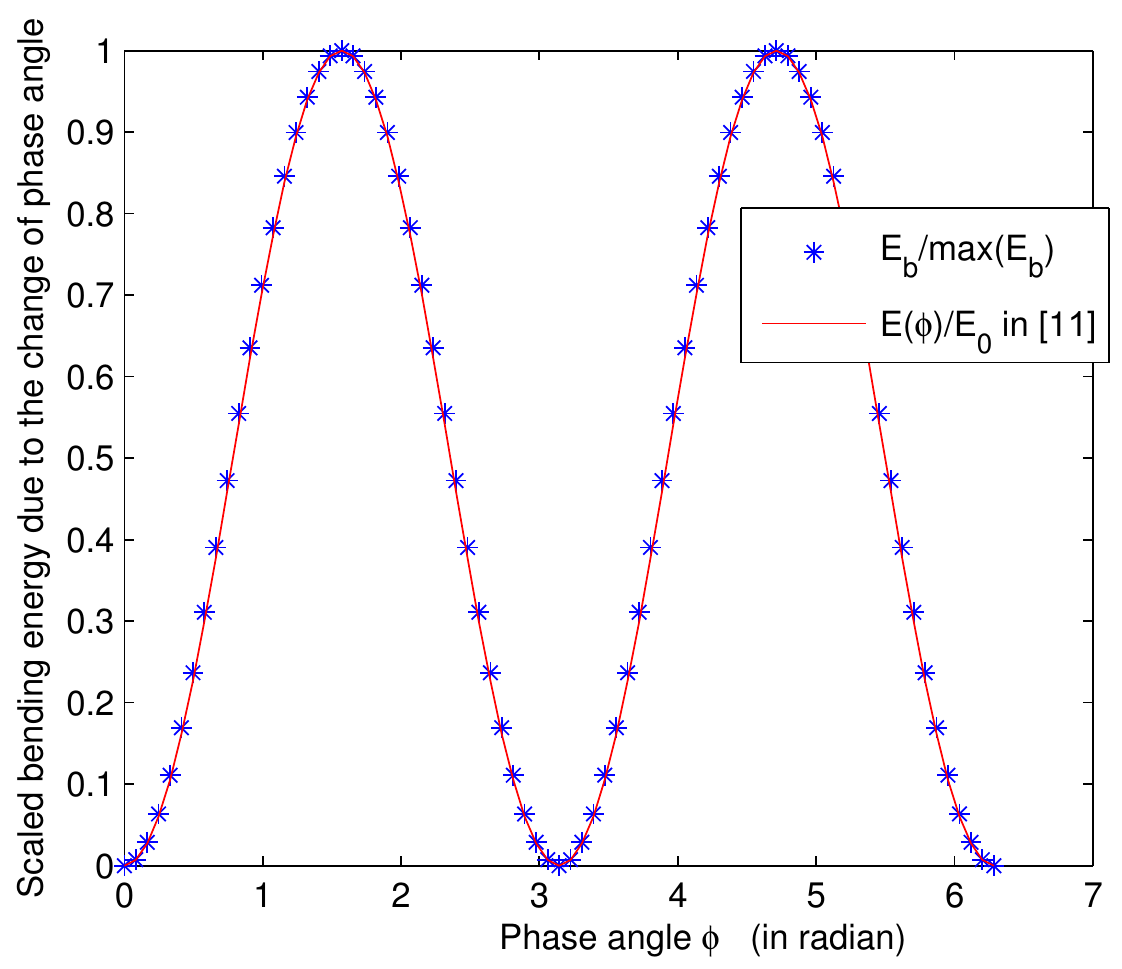} \hskip 10pt
\includegraphics[width=2.25in]{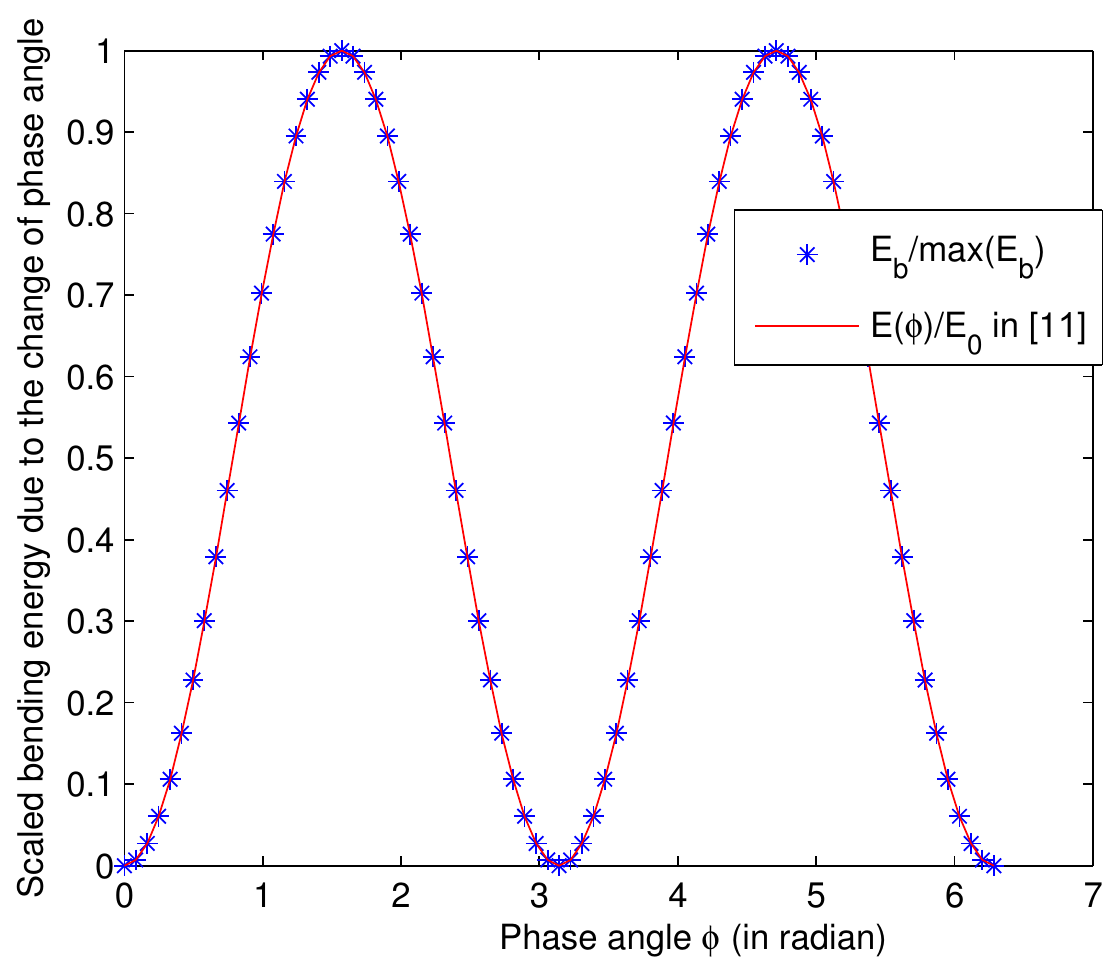} 
\end{center}
\caption{The comparison of the normalized elastic energy term used in ref. [12] (solid line)
and the discrete energy in (\ref{eqn:2}) versus the phase angle $\phi$ for two different 
initial shapes at rest: swelling ratio $s^*$= 0.481 (left) and 0.9 (right).The elastic spring model of the capsule membrane.} \label{FIG.3}
\end{figure}

To obtain a specified initial shape, which also serve as the reference shape at rest  for the natural state,  
we have used the $E_l$ and $E_b$ given in equation (\ref{eqn:2})  with $\theta_i^0=0$ for $i=$1, 
$\dots$, $N$. The capsule is assumed to be a circle of radius $R_0=2.8 \rm{\ \mu m}$ initially and then it
is discretized into $N = 76$ membrane particles so that 76 springs of the same length are formed by 
connecting the neighboring particles. 
The shape change is
stimulated by reducing the total area of the circle through a
penalty function
\begin{equation}
\Gamma_{s}=\frac{k_{s}}{2}(\frac{s-s_{e}}{s_{e}})^{2}\label{eqn:2c}
\end{equation}
where $s$ and $s_{e}$ are the time dependent area of the capsule and the targeted area of the capsule, 
respectively, and the total energy per unit thickness  is modified as $E+\Gamma_{s}$.
Based on the principle of virtual work the force per unit thickness acting on the
$i$th membrane particle now is
\begin{equation}
{\bf F}_{i}=-\frac{\partial(E+\Gamma_{s})}{\partial\label{eqn:2d}
{\bf r}_{i}}
\end{equation}
where ${\bf r}_{i}$ is the position of the $i$th membrane particle.  When the area is reduced, 
each membrane particle moves on the basis of the following equation of motion:
\begin{equation}
m \frac{d^2 {\bf r}_{i}}{dt} +  \gamma \frac{d  {\bf r}_{i} }{dt} ={\bf  F}_i. \label{eqn:2e}
\end{equation}
Here  $m$ and $\gamma$ represent the membrane particle mass and the membrane viscosity of the capsule. 
The position ${\bf r}_i$ of the $i$th membrane particle is solved by discretizing equation (\ref{eqn:2e}) 
via a second order finite difference method. The total energy stored in the membrane decreases as
the time elapses. The final shape of the capsule is obtained as the total energy is minimized and 
such shape is at a stress--free state.
Based on the final shape, as the reference shape (and the initial shape for fluid-capsule
interaction), the angle between two neighboring springs at the $i$th node is $\theta_i^0$
for $i=1$, $\dots$, $N$.  Our bending term in equation (\ref{eqn:2}) is 
similar to the bending energy used by Fedosov  {\it et al.} (e.g., see \cite{Fedosov2010}) which is based on 
a two--dimensional spring network with the natural state as follows
\begin{equation*}
V_{\rm bending}=\displaystyle k_b \sum_{i=1}^{N_s} [1-\cos(\theta_i-\theta_0)] =  k_b \sum_{i=1}^{N_s} 2 \sin^2 (\dfrac{\theta_i-\theta_0}{2}) 
\end{equation*}
where $N_s$ is the total number of the springs, $\theta_i$ is the angle between two triangles sharing 
the $i$th spring as their edges and $\theta_0$ is the angle of uniform natural state.  The term $E_b$ 
defined in equation  (\ref{eqn:2}) 
is a discrete analogue of the elastic energy introduced by Skotheim and Secomb in \cite{Skotheim2007}.  
The pictures of the normalized  $E_b$  and the normalized elastic energy $\sin^2 \phi =0.5-0.5\cos 2\phi$ 
used in  \cite{Skotheim2007}  versus  the phase angle $\phi$ are shown in Fig.  \ref{FIG.3}, in which the graph of the  
normalized term $E_b$  is obtained  by rotating the mass nodes on the membrane of a given fixed shape in a
clockwise manner. Both of them 
have preferred phase angles at zero and 180 degrees.
We also modify the bending energy per  unit thickness  to a weighted sum of both uniform and  nonuniform natural states
to have a weaker effect of the nonuniform natural state:
\begin{equation}
E_b=\frac{k_b}{2}\left(\left(1-\alpha\right)\sum_{i=1}^{N}\tan^2\left(\frac{\theta_i}{2}\right)+
\alpha\sum_{i=1}^{N}\tan^2\left(\frac{\theta_i-\theta_i^0}{2}\right)\right).\label{eqn:2b}
\end{equation}
Here, $\alpha$ indicates the weight of the nonuniform natural state. 

 
%
%

In equations (\ref{eqn:2}), (\ref{eqn:2c}) and (\ref{eqn:2b}), $E_{l}$, $E_{b}$, and $\Gamma_{s}$ 
represent energies [N m] per unit thickness of 1 m, and thus the units 
of $k_{l}$, $k_{b}$, and $k_{s}$ are Newton [N].  In this paper, we only consider the capillary 
number instead of shear rate, because it is the key number to determine the behavior of capsules 
in two--dimensional flow. The value of the swelling ratio of a capsule is $s^{*}=s_e/(\pi R_0^2)$.
The values of parameters for modeling  capsules are as follows:
The spring constant is $k_{l}=5\times10^{-8}$ $\rm{N}$, the penalty coefficient is $k_s= 10^{-5}$ $\rm{N}$,
and the  bending constant is $k_{b}=5\times 10^{-10}$ $\rm{N}$. 
Using the above chosen parameter values, 
the area of the initial shape of the capsule has less than 0.001\% difference from the given equilibrium area
$s_e$, and the length of the capsule perimeter of the initial shape  has less
than 0.005\% difference from the circumference of the initial circle.

\subsection{Computational methods}

In simulations, the capsule with the aforementioned initial shape is suspended in a fluid which has a 
density $\rho=1.00$ $\rm{g/cm^{3}}$ and  a dynamical viscosity  $\mu$ = 0.012 $\rm{g/(cm \ s)}$. 
The viscosity ratio of the inner and outer fluid of the capsule membrane is fixed at 1.0. 
The motion of a capsule in fluid flow is simulated by combining the immersed boundary
method \cite{Peskin1977, Peskin1980, Peskin2002} and the elastic spring model given in equation 
(\ref{eqn:2b}) for the membrane.  The Navier--Stokes equations for fluid flow have been solved 
by using an operator splitting technique and finite element method
with a structured triangular mesh so that the specialized fast solver, such as FISHPAK by
Adams et al. \cite{Adams}, can be used to solve the fluid flow.  The details of computational methodologies 
are discussed in \cite{Shi2012IJNMF}.


\section{Validation}
\subsection{An inextensible capsule tank--treading in the shear flow}
\begin{figure}
\begin{center}
\leavevmode
\includegraphics[width=2in]{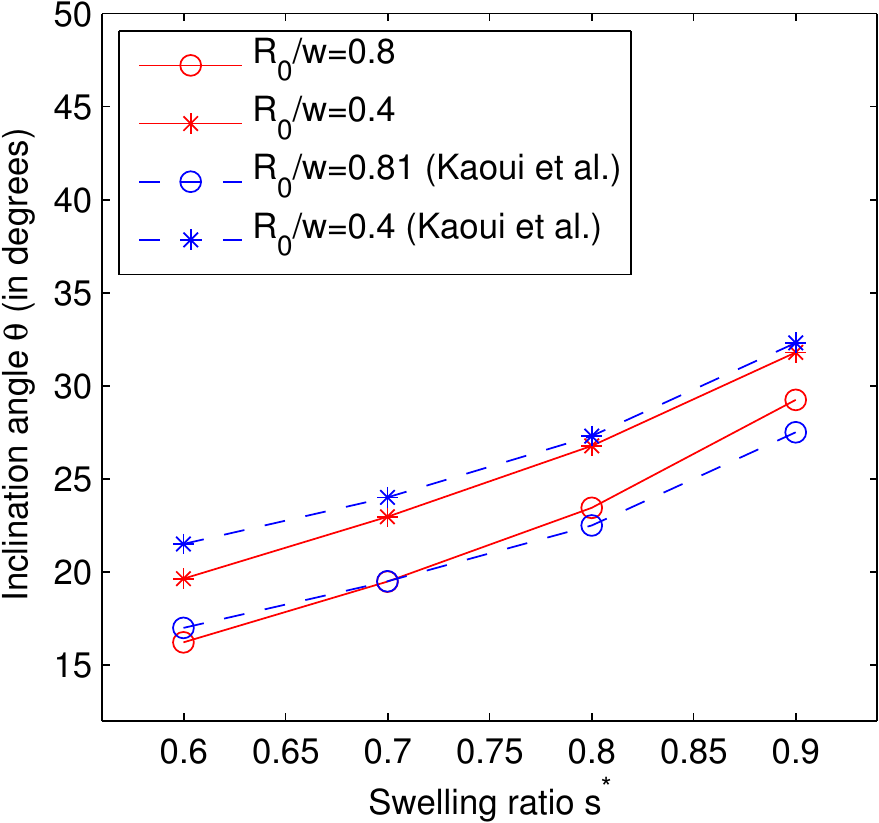}
\includegraphics[width=2in]{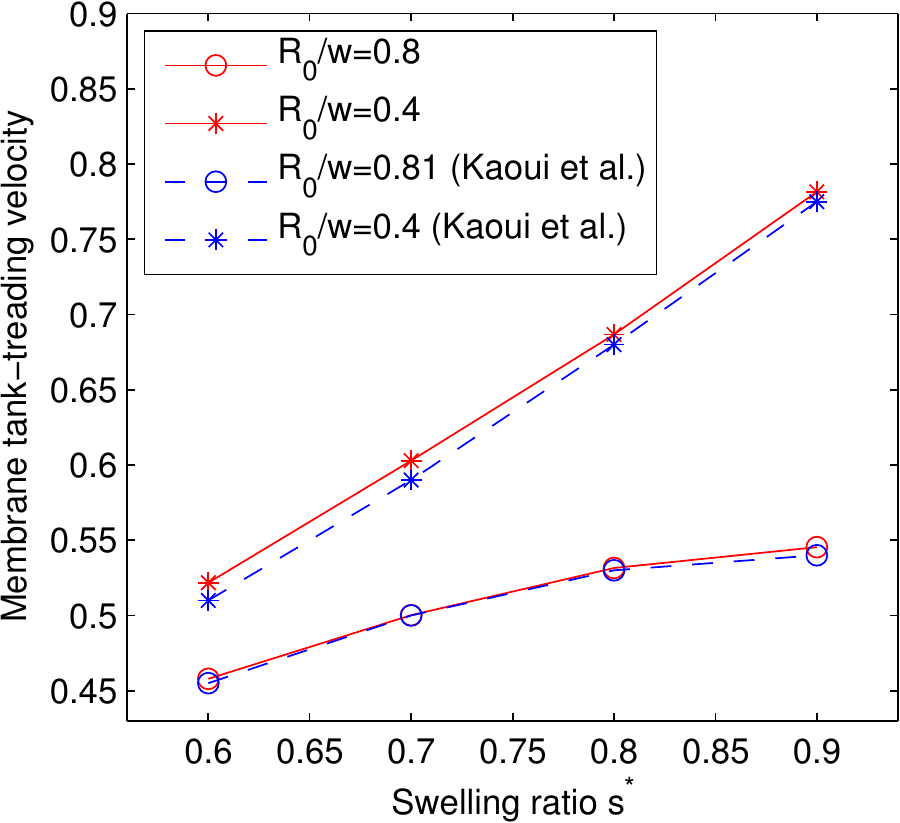}
\end{center}
\caption{(Color online). Steady inclination angle as a function of the capsule swelling ratio for two degrees of
confinement $R_0/w$= 0.4 and 0.8.} \label{FIG.4}
\end{figure}

First, we present the simulating results of an inextensible capsule with uniform nature state
(i.e., $\alpha=0$ in equation (\ref{eqn:2b}) and $\theta_i^0$=0, for $i=1$, $\dots$, $N$) suspended in a 
linear shear flow with shear rate $\gamma=500\rm{/s}$. The dimensions of the computational domain are
$100\rm{\mu m }\times 7\rm{\mu m}$ and $100\rm{\mu m} \times 14\rm{\mu m}$.
The two degrees of confinement are 0.8 for the narrower domain and 0.4 for the wider domain,
respectively. The grid resolution for the computational domain is 80 grid points per 10$\rm{\mu m}$.
The time step $\Delta t$ is $1 \times 10^{-4} \rm{ms}$. The initial velocity of the fluid flow is
zero everywhere and the initial positions of the mass center of the capsule are the center of both domains. 
In both wider and narrower domains,  the capsule performs a steady tank--treading motion with a fixed inclination angle
which depends on the swelling ratio $s^*$ and degree of confinement.
The steady inclination angles and the membrane tank--treading velocities of two different degrees of confinement for
five values $s^*$= 0.6, 0.7, 0.8 and 0.9 are presented in Figure \ref{FIG.4},
which show good agreement with the results by Kaoui {\it et al.} \cite{Kaoui2011}. The inclination angle increases monotonically for both two degrees of
confinement with increasing the value of the swelling ratio $s^*$. For the same swelling ratio, the
bigger is the degree of confinement, the smaller is the  steady inclination angle. 
On the other side, tank--treading velocity increases almost linearly with respect to increasing of 
swelling ratio in the narrower channel, while in the wider domain
the increasing of tank--treading velocity has slowed down until it arrives almost maximum around $s^*\sim 0.9$.
The same qualitative tendency is given in \cite{Kaoui2011}-\cite{Beaucourt}.
We also keep track of the area and the perimeter of the capsule during the simulations. 
The variation is less than $0.1 \%$ in the area and $ 0.5 \%$ in the perimeter.
Our model does keep capsules as an inextensible one in simulations.

\subsection{A go--and--stop experiment for the shape recovery}

\begin{figure}[!htp]
\begin{center}
\includegraphics[width=5in]{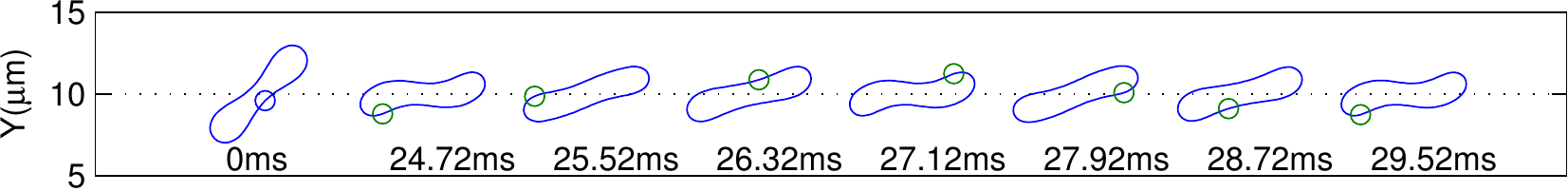}\\
\includegraphics[width=5in]{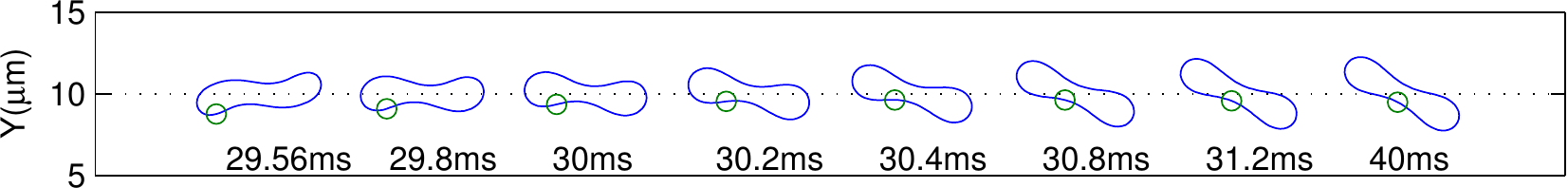}\\
\includegraphics[width=5in]{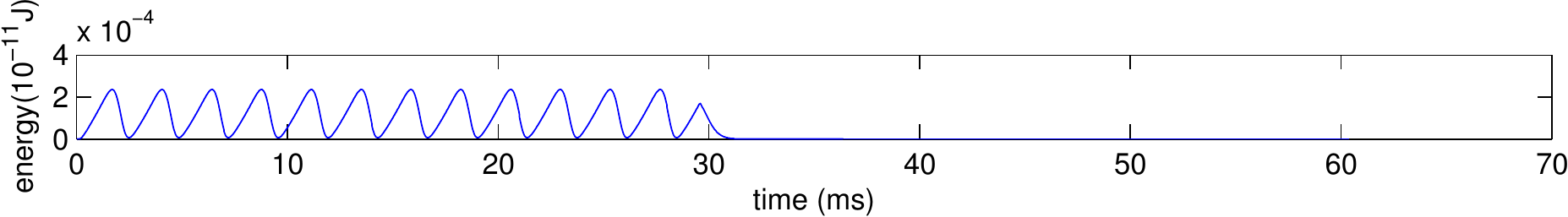}
\end{center}
\caption{(Color online) A go--and--stop experiment: the snapshots (top and middle) and history of energy (bottom) over the unit thickness
of the capsule in shear flow at $C_a=6.365$. The moving walls are stopped at time 29.52 ms.}\label{FIG.5}
\end{figure}

A capsule or cell is said to have the shape memory if after stopping the flow, the deformed 
one will go back to its initial shape with any part of the
membrane regaining its original position, i.e., the rim always returns to
the rim and the dimple is always back to the dimple.
To show that the capsule with a nonuniform natural state modeled by (\ref{eqn:2})--(\ref{eqn:2d}) 
does have the shape memory property,  we have simulated several cases of the go--and--stop.

\begin{figure}[!htp]
\begin{center}
\includegraphics[width=5in]{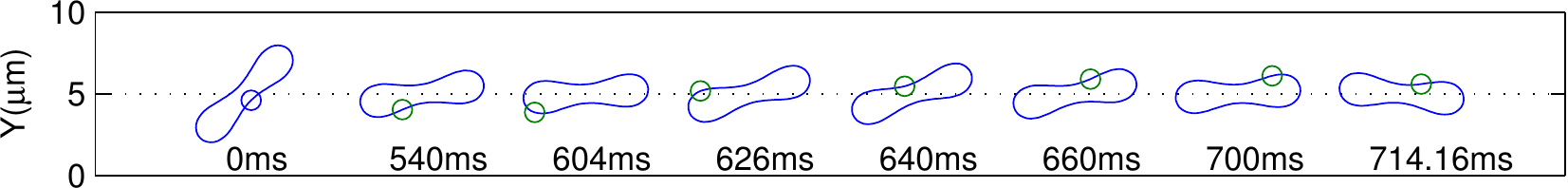}\\
\includegraphics[width=5in]{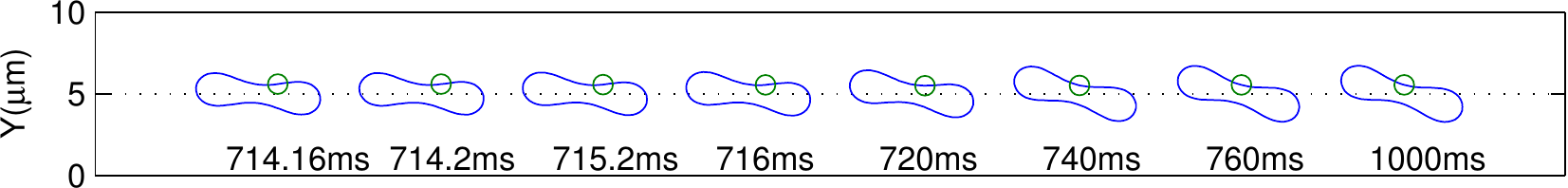}
\end{center}
\caption{(Color online) Snapshots of the capsule in shear flow with $C_a=0.182$ in narrower channel for $\alpha=0.05$. 
At time $t=714.16$ ms, the motion of two walls is stopped.}\label{FIG.6}
\end{figure}

\begin{figure}[!htp]
\begin{center}
\includegraphics[width=5in]{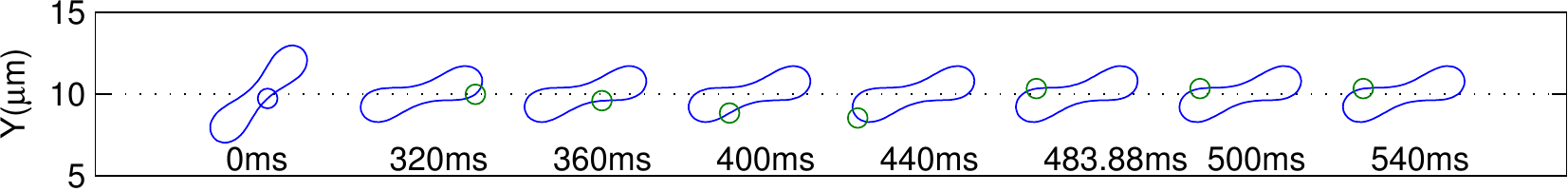}
\end{center}
\caption{(Color online) Snapshots of the capsule in shear flow with $C_a=0.091$ in a wide channel for $\alpha=0$. 
At time $t=483.88$ ms, the motion of two walls is stopped.}\label{FIG.7}
\end{figure}

Figure \ref{FIG.5} shows a capsule with $\alpha=1$ (fully non--uniform) of the swelling ratio $s^{*}=0.481$ suspended in a simple 
shear flow at the capillary number $C_a=6.365$ in a channel of the length $80\mu$m and width $20\mu$m. The top and bottom 
walls of the channel are driven at the same speed in opposite directions as shown in Figure \ref{FIG.1.1}. The capsule is at 
its reference shape at time $t=0$ s, with a membrane particle at the dimple marked by a small ``o''.
At the beginning, the capsule performs a tank--treading motion with a swinging mode
and the marked position tank--treads along the membrane as in Figure \ref{FIG.5} (top). We then
suddenly stop the motion of two walls by setting the boundary conditions on  them
equal to zero, and observe (in Figure \ref{FIG.5} (middle)) that the capsule first returns to the biconcave shape 
very quickly and then the marked position tank--treads back to its initial location
on the membrane, this behavior is first observed in experiments by T. M. Fischer in \cite{Fischer2004}. 
The capsule energy, plotted in Figure \ref{FIG.5} (bottom), is minimized when the capsule is 
at its natural state. During TT with a swinging mode, the capsule energy changes periodically with 
the period equal to the half of the period of oscillation due to the symmetry of the capsule 
natural state. After stopping the motion of two walls, the capsule energy returns to the one at its natural state. 
At lower $\alpha$ values, we get the same behavior of the rim back to the rim and the dimple back to the dimple 
after the walls stopped moving, as shown in Figure \ref{FIG.6} for $\alpha=0.05$.
For $\alpha=0$ (see Figure \ref{FIG.7}), the tank--treading motion stops right after the wall motion stops; but
the membrane particle from dimple does not go back to the dimple. This suggests that  with
$0 < \alpha \le 1$, we can model the behavior of the shape memory.
Also  with larger $\alpha$ value, we get stronger memory of the reference shape in the sense that membrane tank--treads back to 
the reference shape faster.  


\section{Tumbling and tank-treading with a swinging mode}

We have  considered a capsule with a fully non--uniform natural state ($\alpha=1$) suspended in shear 
flow at the capillary numbers $C_a=$ 0.455, 9.093, and 90.934 in a channel of the 
length $80\mu$m and width $20\mu$m. These  capillary numbers give rise to two typical 
motions of the capsule, namely, (i) a tumbling motion and 
(ii) a tank--treading motion with a swinging mode, respectively.  
The inclination angle $\theta$ and the phase angle $\phi$ defined in Figure \ref{FIG.1.2} and Figure \ref{FIG.1.3}, 
respectively, are used to described the capsule motion.
Since the shape of capsule in shear flow is symmetric about its mass center and so is the reference 
angles in the bending term $E_b$, we can restrict the range of both inclination and phase angles to 
$[-90^{\circ},90^{\circ}]$. If the inclination angle has 
the local maxima and  minima within $(-90^{\circ},90^{\circ})$, the capsule is having a swinging mode; 
otherwise  if the inclination angle decreases monotonously to $-90^{\circ}$ then jump to $90^{\circ}$, it has 
a tumbling motion.  The criteria for the phase angle are the opposite. When the local maxima and minima 
of the phase angle are in $(-90^{\circ},90^{\circ})$, the capsule has the tumbling motion. If
the phase angle increases monotonously to $90^{\circ}$  and then jumps to $-90^{\circ}$, it means 
that the capsule is   tank--treading. 

\begin{figure}[!tp]
\begin{center}
\includegraphics[width=4.7in]{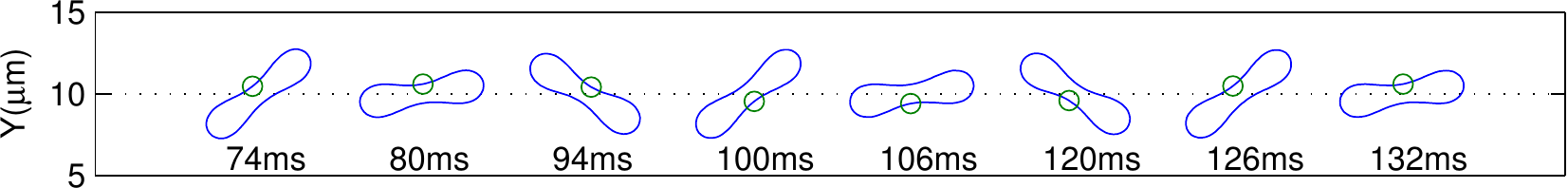}\\
\includegraphics[width=4.7in]{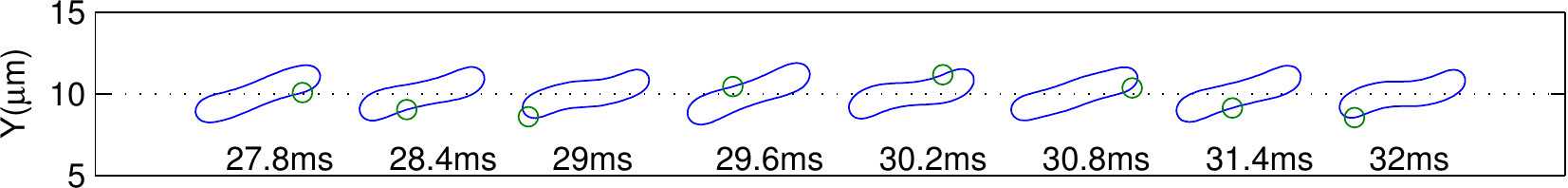}\\
\includegraphics[width=4.7in]{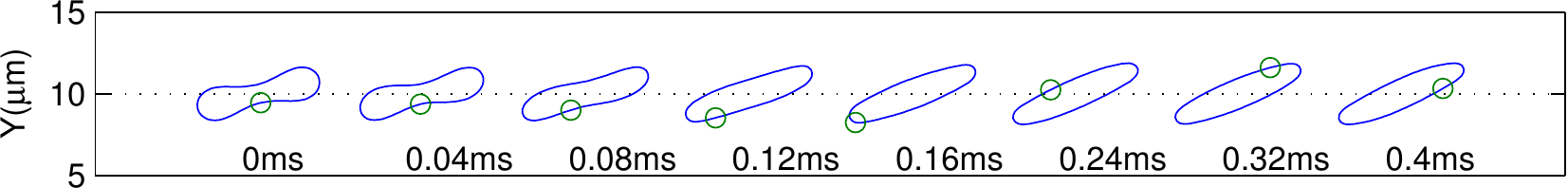}
\end{center}
\caption{(Color online) Snapshots of the capsule shape and orientation at the capillary number 
$C_a$= 0.455 (top), 9.093 (middle), and 90.934 (bottom)
illustrate the tumbling (top) and tank--treading with swinging mode (middle and bottom), respectively.}\label{FIG.8}
\end{figure}
\begin{figure}[ht]
\begin{center}
\includegraphics[width=2.35in]{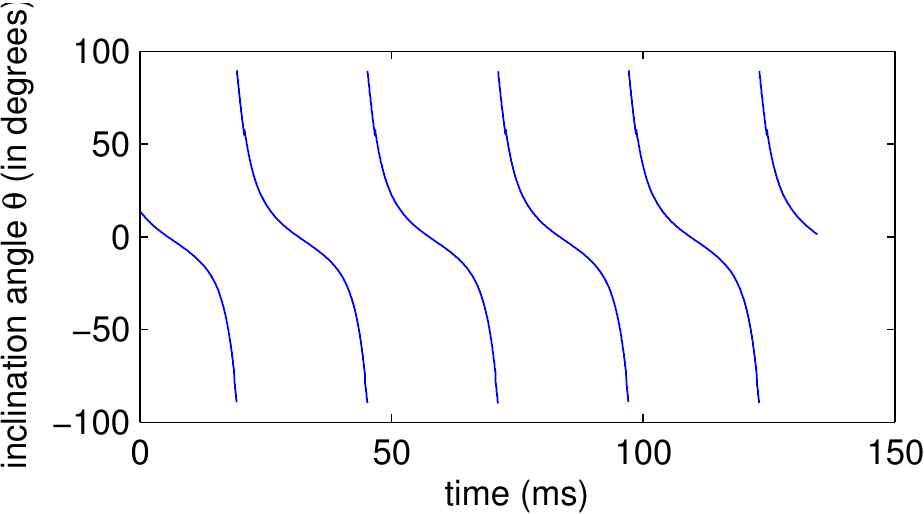}
\includegraphics[width=2.35in]{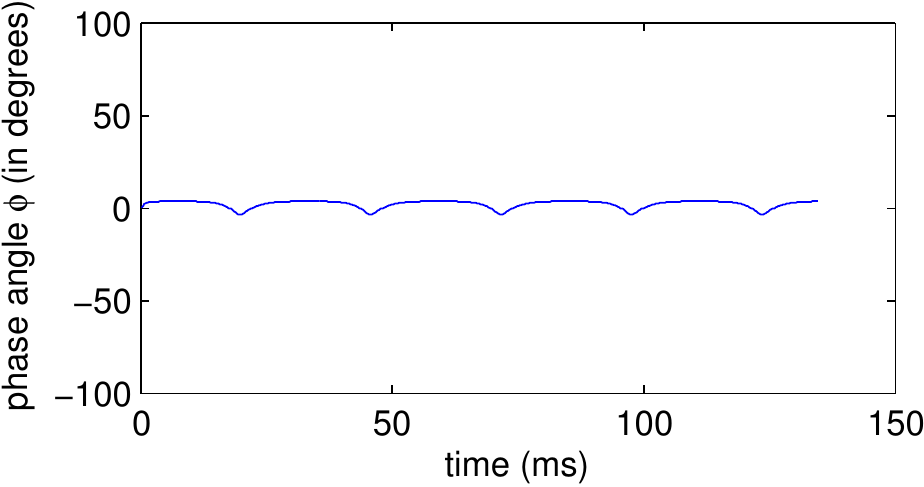}\\
\includegraphics[width=2.35in]{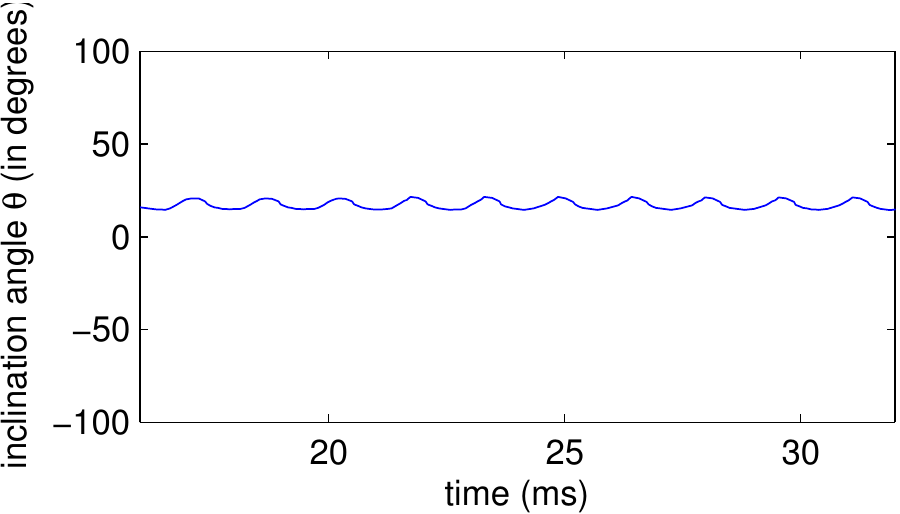}
\includegraphics[width=2.35in]{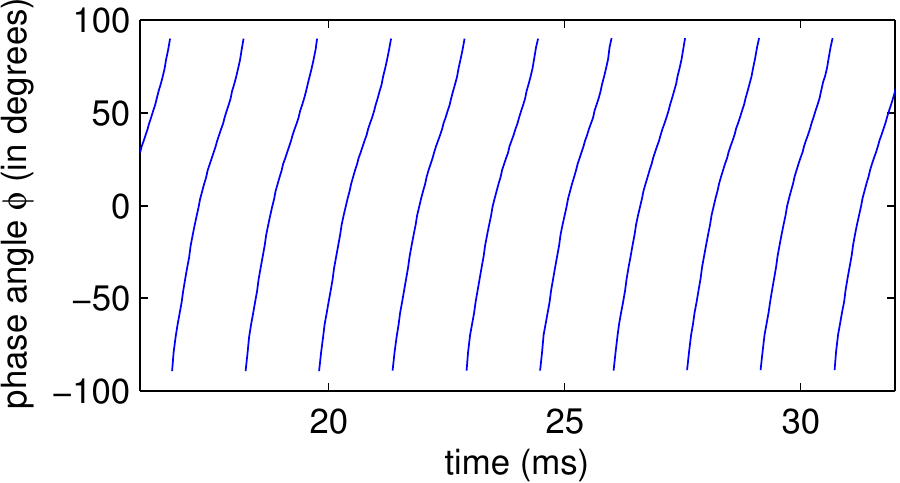}\\
\includegraphics[width=2.35in]{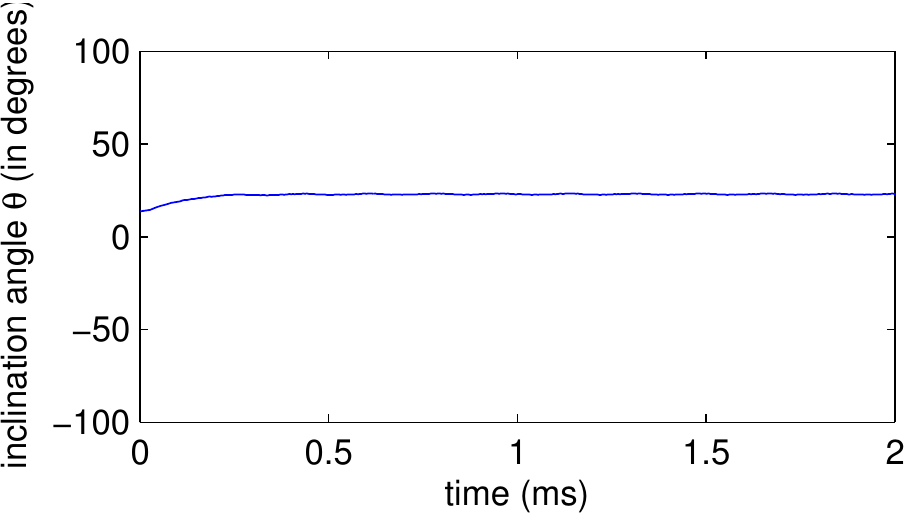}
\includegraphics[width=2.35in]{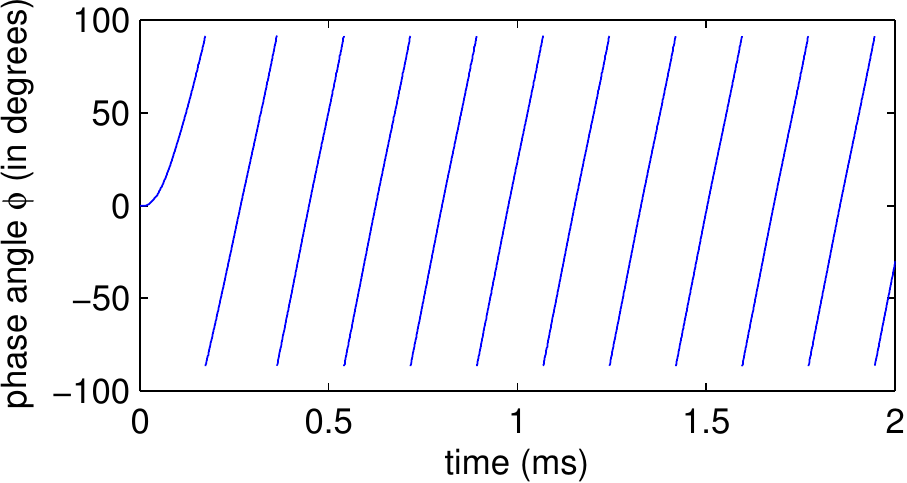}
\end{center}
\caption{(Color online) Histories of the inclination angle (left) and the phase angle (right) at the capillary number 
$Ca$= 0.455 (top), 9.093 (middle), and 90.934 (bottom) associated with the
tumbling (top) and tank--treading with swinging mode (middle and bottom), respectively.}\label{FIG.9}
\end{figure}

\begin{figure}[!ht]
\begin{center}
\includegraphics[width=4.7in]{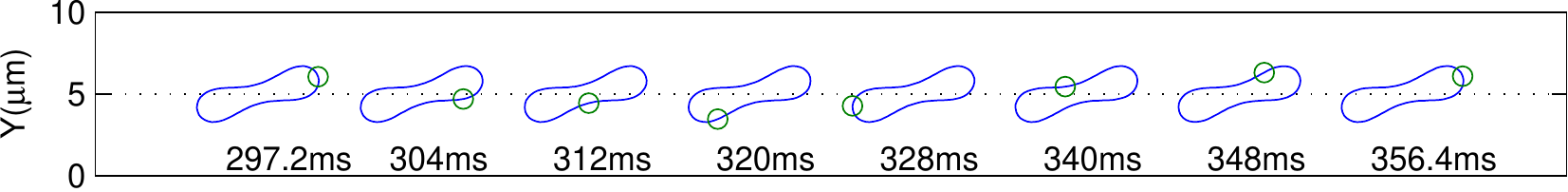}\\
\includegraphics[width=4.7in]{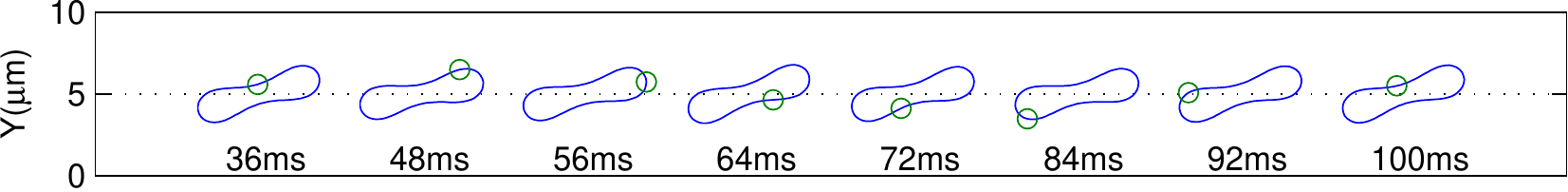}\\
\includegraphics[width=4.7in]{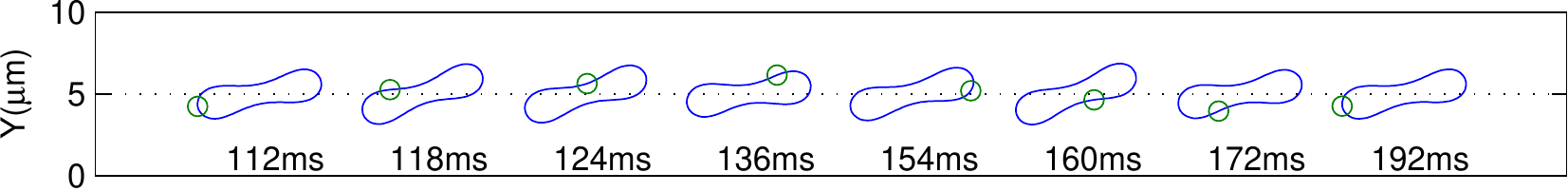}\\
\includegraphics[width=4.7in]{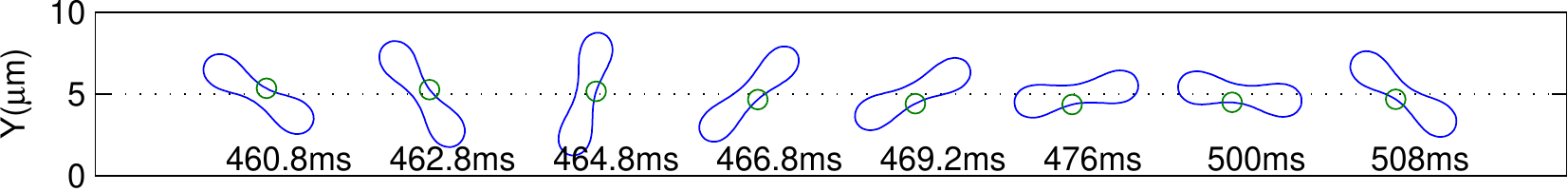}\\
\includegraphics[width=4.7in]{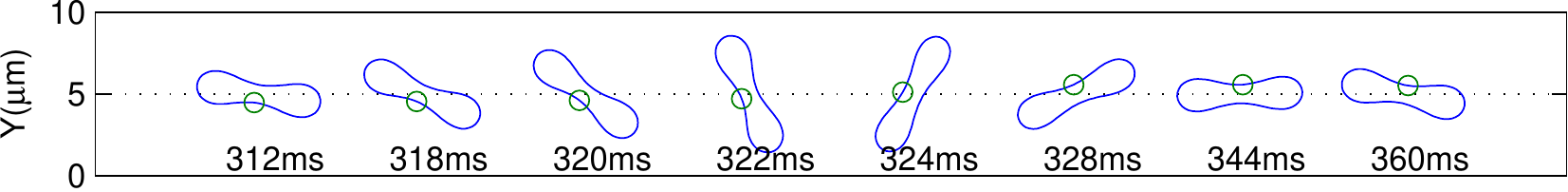}\\
\end{center}
\caption{(Color online)  Snapshots of the capsule motion of $s^{*}$ = $0.481$ with different 
nonuniform natural states at $\alpha$=0, 0.05, 0.1, 0.5 and 1, respectively, (from top to bottom)
$C_a$= 0.455.}\label{FIG.10}
\end{figure}

At the capillary number $C_a=$ 90.934, 
the capsule is elongated to an elliptical shape and performs a tank--treading motion with almost the
same shape and inclination angle. When reducing the capillary number to 9.093,
the capsule tank--treading with periodically oscillating inclination angle 
is obtained. We also observe that the shape of the capsule is deformed periodically. 
And at $C_a=$ 0.455, the capsule keeps tumbling periodically with a small oscillation of the phase angle. 
These motions are illustrated by 
the snapshots and histories of the inclination and phase angles of the capsule in 
Figures \ref{FIG.8} and \ref{FIG.9}.  During the tank--treading  with a swinging mode, the capsule 
keeps changing its shape periodically with respect to the period of tank--treading. 
The shape deformation comes from the fact that during these motions 
the membrane tends to reduce the difference between the current
bending angle $\theta_i$ and the reference angle $\theta_i^0$ to
minimize the elastic energy. The obtained swinging mode is
always coupled with the tank--treading motion. At the higher shear rate 
of $C_a=$ 90.934, tank--treading motion is still accompanied by a
swinging mode where the oscillation of inclination angle is very small (see Figure \ref{FIG.9}). 
To demonstrate that the effect of the bending  term $E_b$ 
given in equations (\ref{eqn:2}) and (\ref{eqn:2b}) is also a key factor (besides the viscosity ratio) on the 
capsule motion in shear flow, we have presented in Figure \ref{FIG.10} that the capsule at different nonuniform 
natural state undergoes either tank--treading motion or tumbling  at a shear rate 500 s$^{-1}$. As the effect of
the nonuniform natural state is weaker, the capsule just tank-treads; but under stronger effect, the capsule
does tumble as shown in Figure \ref{FIG.10}. We have also obtained that,
for all choices of capillary numbers, the capsule with the uniform natural 
state undergoes tank--treading motion, which is same as  in \cite{Tsubota2010}.

\begin{figure}[!tp]
\begin{center}
\includegraphics[width=2.5in]{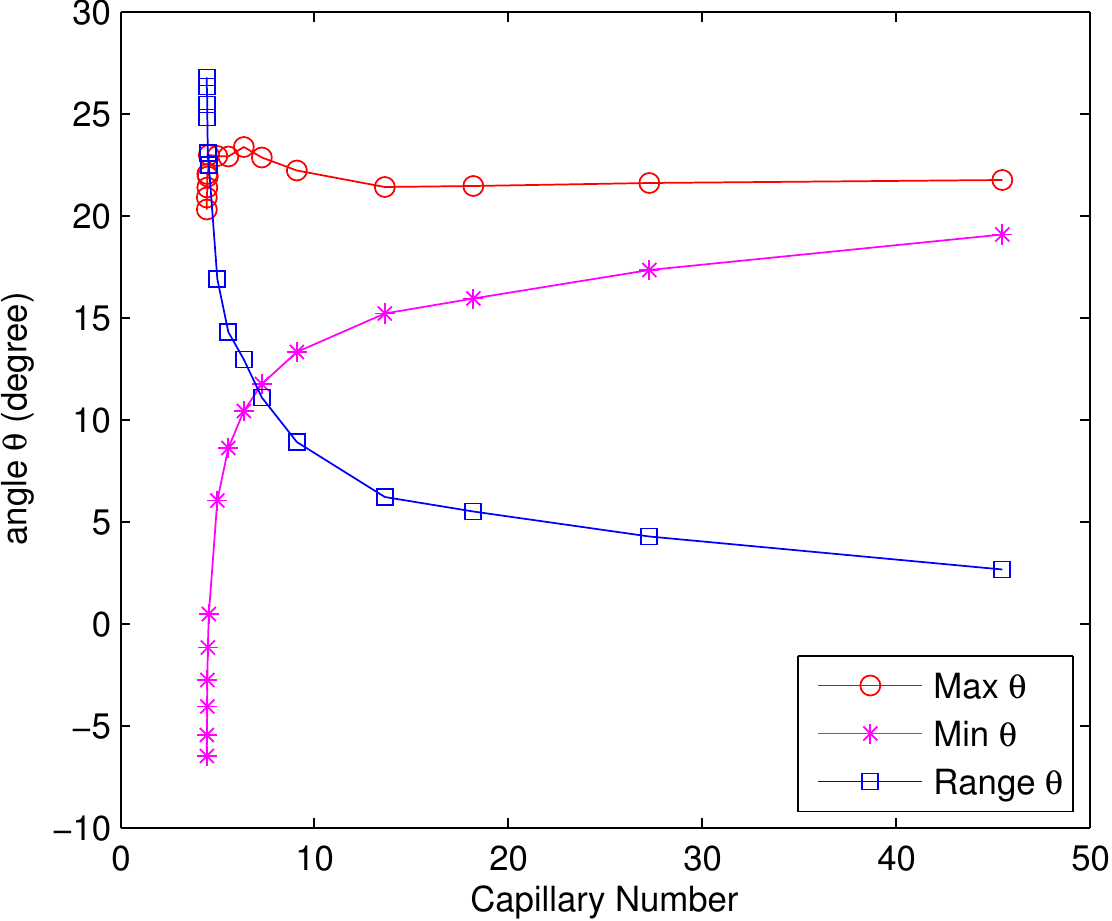}
\includegraphics[width=2.5in]{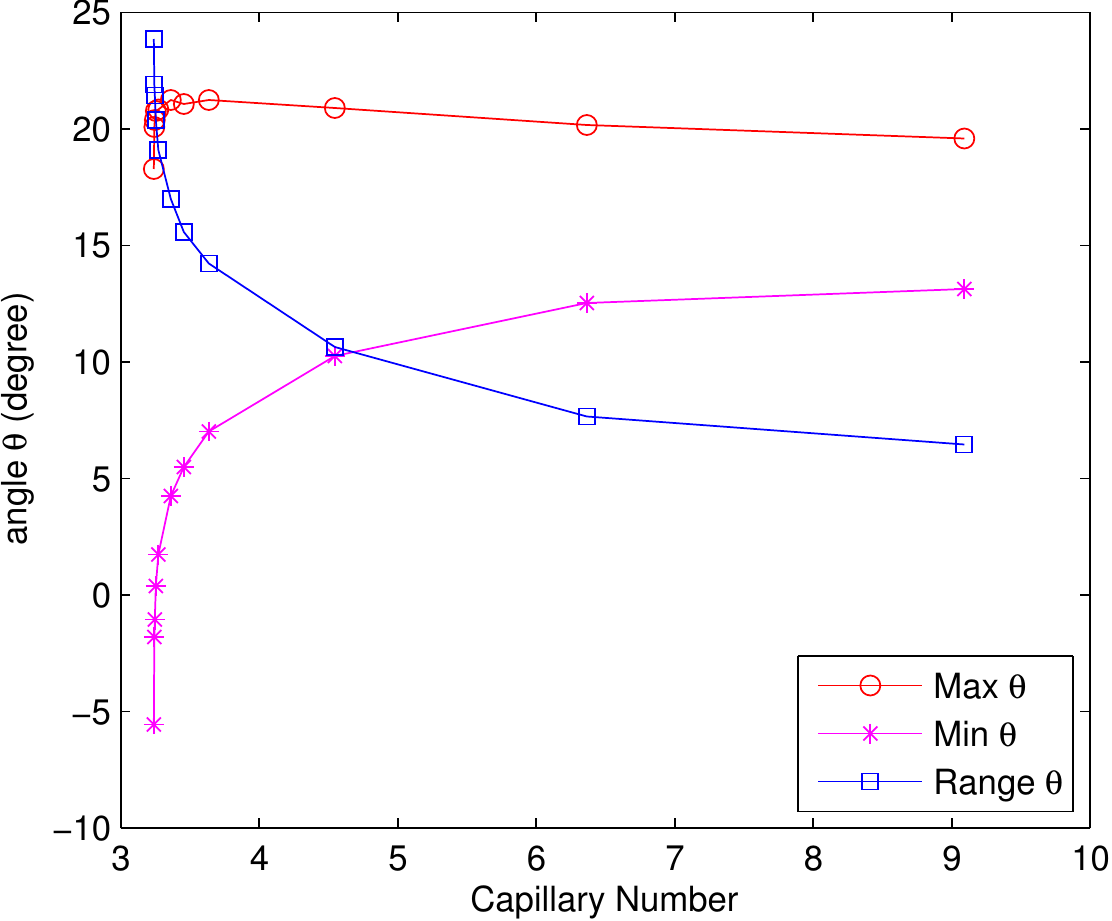}
\end{center}
\caption{(Color online) Inclination angles in wider channel (left) and narrower
channel (right) with respect to capillary numbers.}\label{FIG.11}
\begin{center}
\includegraphics[width=2.5in]{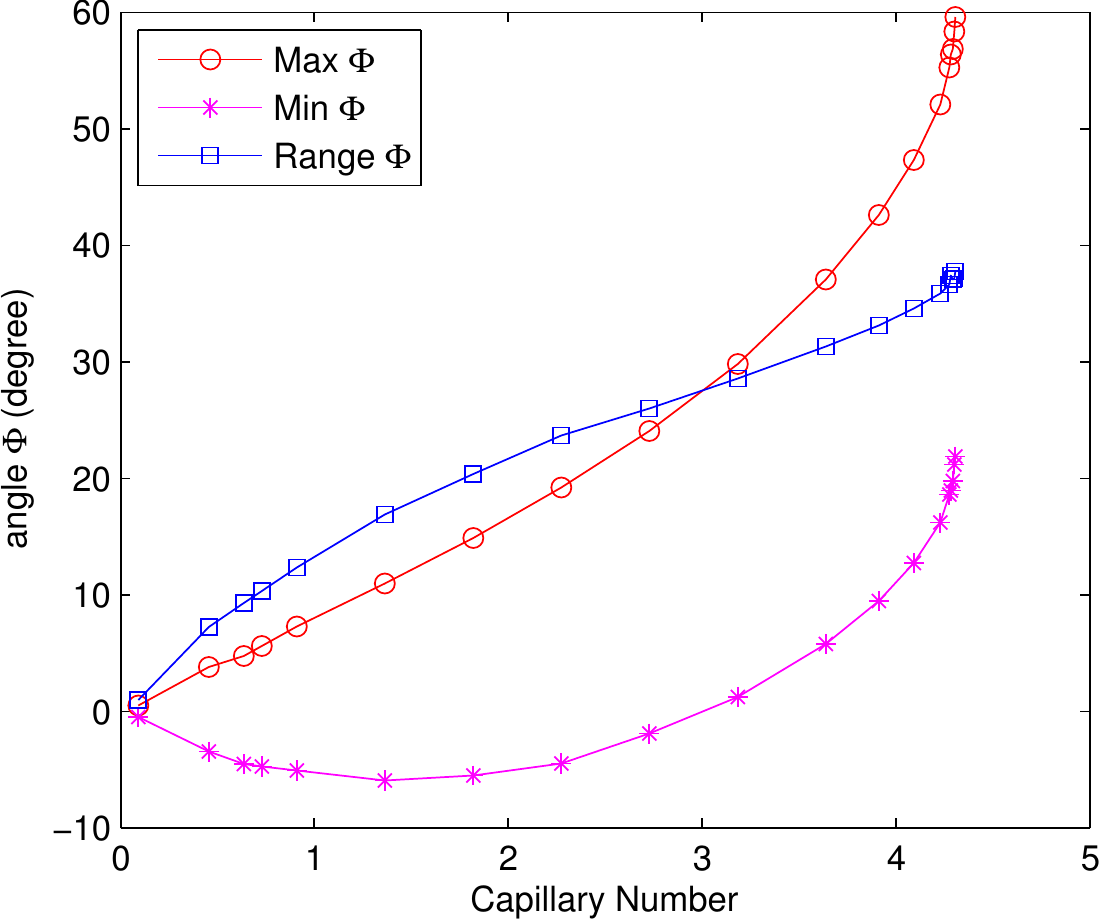}
\includegraphics[width=2.5in]{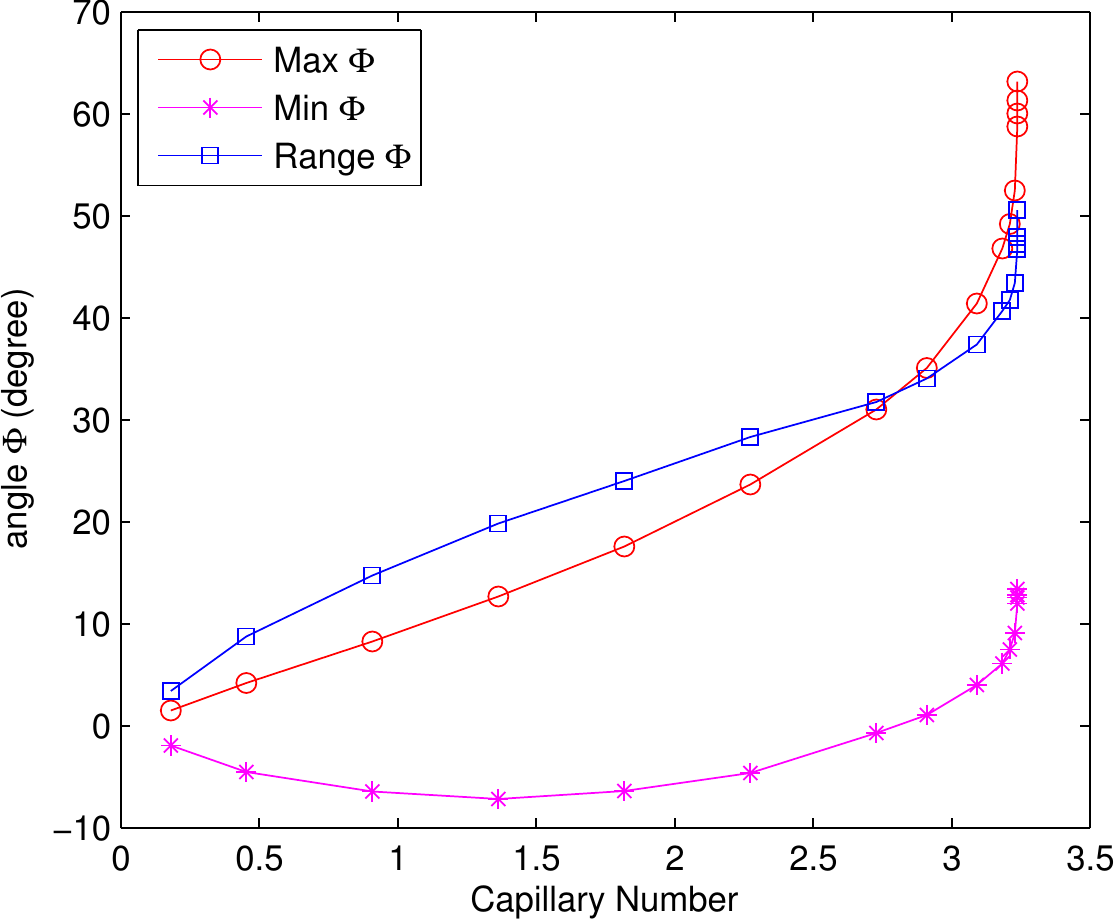}
\end{center}
\caption{(Color online) Phase angles in wider channel (left) and narrower
channel (right) with respect to capillary numbers.}\label{FIG.12}
\end{figure}

The inclination and phase angles in Figures \ref{FIG.11} and \ref{FIG.12} do behave similarly to those in 
\cite{Skotheim2007}. While the capsule is in swinging mode, the range of inclination angle decreases as the capillary 
number is increasing. Also a larger confinement ratio results in the speeding up of the rate of decreasing. 
On the other hand, while the capsule tumbles, the range of phase angle increases with the increasing of capillary number. 
but it does not show much dependence on the confinement ratio.  The range of phase angle shows that the capsule tank-treads
backward and forward while tumbling.  

\section{The intermittent region between tumbling and tank-treading}

\begin{figure}[ht]
\begin{center}
\includegraphics[width=5.in]{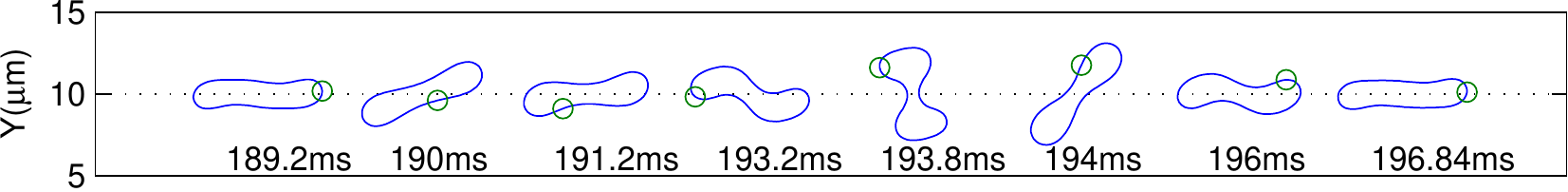}\\
\includegraphics[width=2.5in]{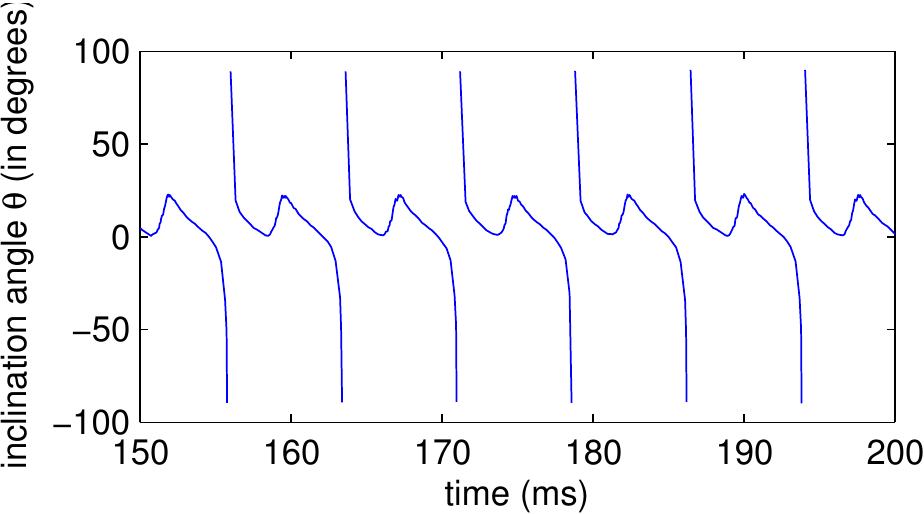}
\includegraphics[width=2.5in]{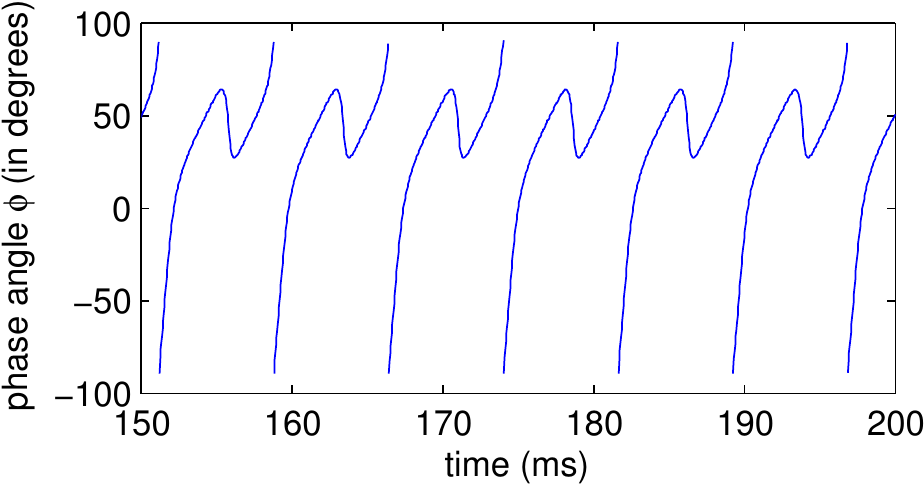}\\
\includegraphics[width=5.in]{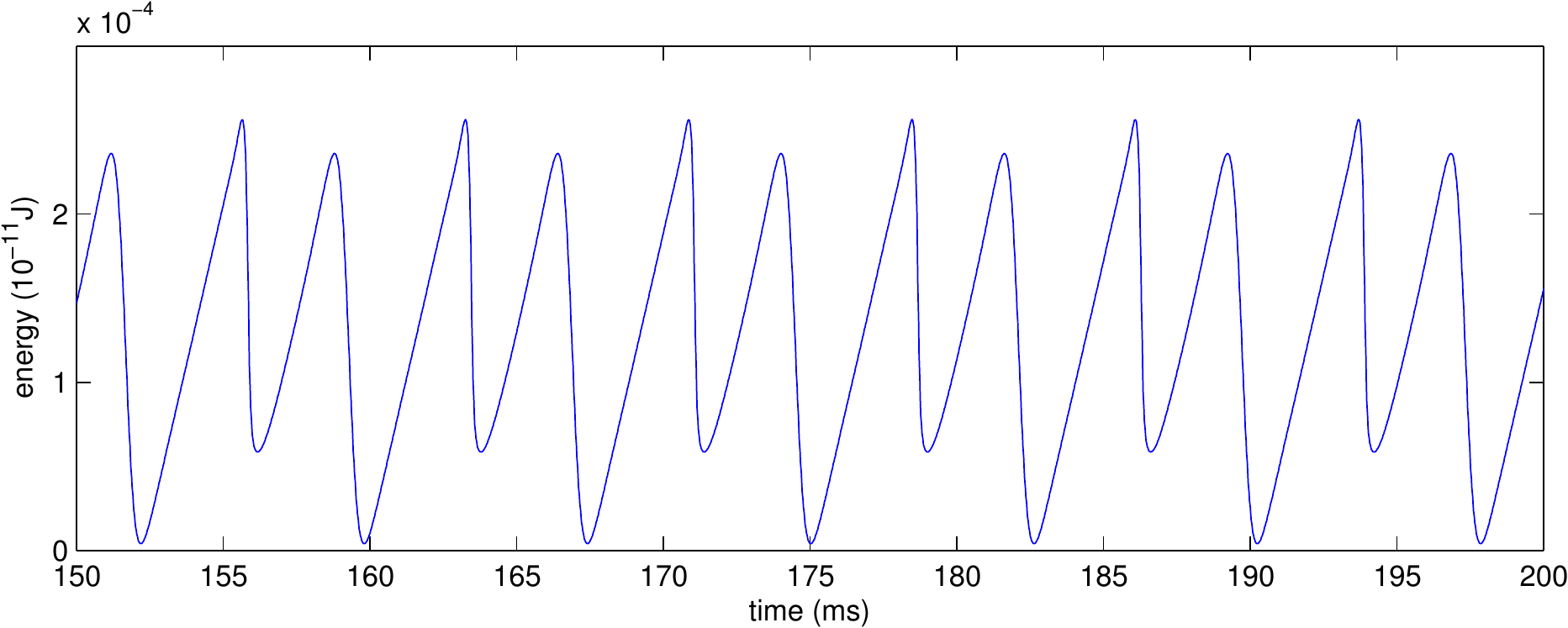}
\end{center}
\caption{(Color online) Snapshots of the capsule shape and orientation in one cycle (top), histories of the inclination 
angle (middle left), the phase angle (middle right), and the energy over the unit thickness (bottom) of the capsule at
$Ca$= 4.410.}\label{FIG.13}
\end{figure}

\begin{figure}[!tp]
\begin{center}
\includegraphics[width=2.5in]{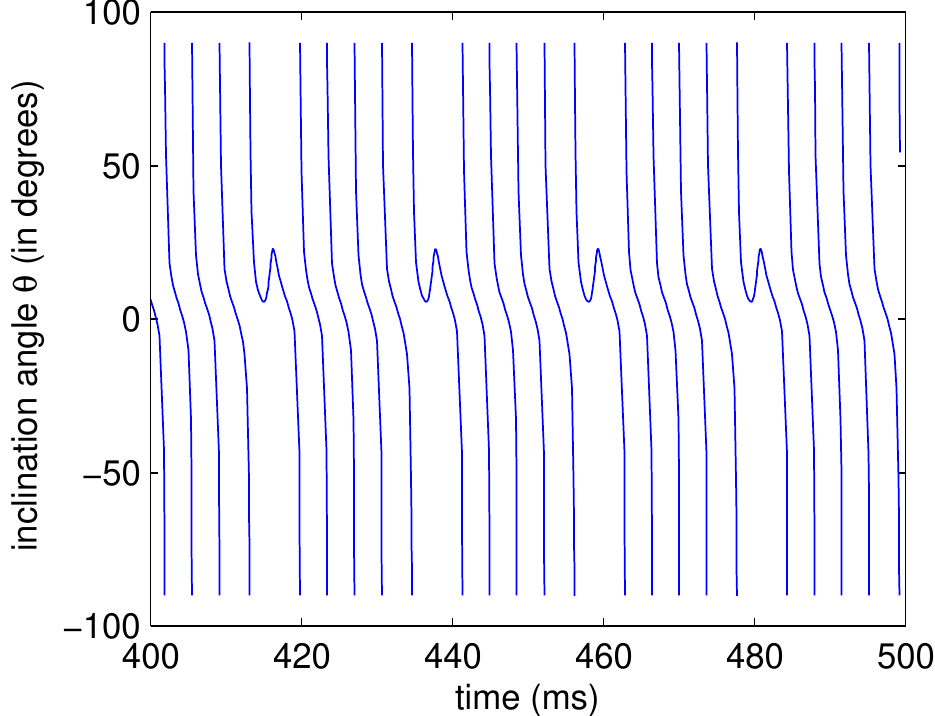}
\includegraphics[width=2.5in]{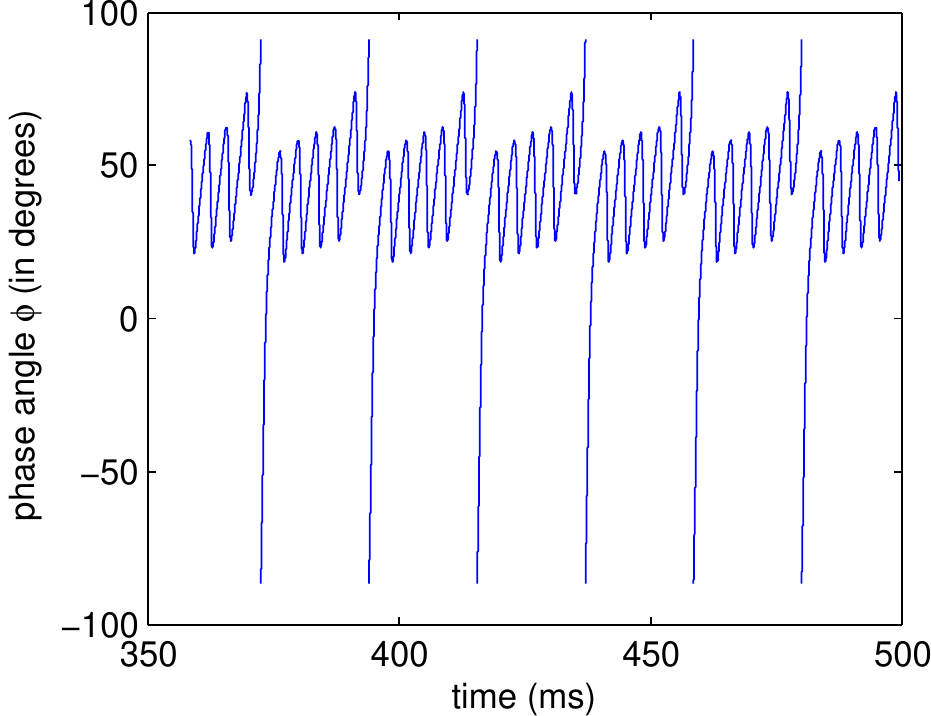}
\end{center}
\caption{(Color online) Histories of the inclination angle (left) and the phase angle (right) of the capsule at
$Ca$= 4.319}\label{FIG.14}
\end{figure}

The intermittent behavior of capsule motion has been  obtained  in \cite{Skotheim2007} due to the elastic energy 
term. Our computational results show that the bending energy density $E_b$ in  equations (\ref{eqn:2}) and  
(\ref{eqn:2b}) does play a similar role  and gives rise to the intermittent behavior of the capsule like those 
obtained in \cite{Skotheim2007}, a mixture of tumbling and TT with a swinging mode. When having the mixture of two 
motions, we notice that (i) while tumbling, the membrane does tank-tread backward and forward within a small range as 
presented in Figure \ref{FIG.12} and (ii)  during tank-treading with a swinging mode, the inclination angle of the 
capsule only oscillates within a small range as in Figure \ref{FIG.11}.
An example of a mixed dynamics of the capsule in the intermittent region obtained  at $C_a=4.410$ with  $\alpha=1$ 
is that  the capsule performs one tumbling and one  TT with a swinging mode in each cycle in a wider channel.  
Figure \ref{FIG.13} shows the snapshots of the capsule motion in one cycle
as well as the histories of the inclination angle, the phase angle and the capsule energy. 
Since the capsule performs one tumbling and one TT with a swinging mode in each cycle, both angles show jumps 
(see the middle two of Figure \ref{FIG.13}). From the snapshots of the capsule motion (the top one of Figure \ref{FIG.13}), the marker ``$\circ$'' on the membrane moves in the clockwise direction at the first 4 
different times while oscillating, then moves backward and forward, while tumbling, as shown at $t=193.8$ ms
194 ms, and 196 ms, and finally finishes one complete tank-treading at $t=196.84$ ms. 
Such motion shows a mixed dynamics of tumbling and one TT with a swinging mode in each cycle.
The energy required for the tumbling in each cycle appears 
to be higher than that during the TT with a swinging mode, because 
when the capsule oscillates, it has an elongated shape which is much closer to the original
biconcave shape than those shapes when the capsule tumbles.
Another typical example of the mixed dynamics of the 
capsule  with $\alpha=1$  at $C_a=4.4319$ in the intermittent region  is shown in  Figure \ref{FIG.14}.   
We have observed that the capsule has  five tumbling  and one TT with a swinging mode in each cycle 
(i.e., a periodic pattern of the mixed dynamics). In  Figure \ref{FIG.14}, the membrane moves backward 
and forward slightly during each tumbling; it finally moves across the hurdle and finishes a period of
a tank-treading.

\begin{figure}[!htp]
\begin{center}
\subfloat[][]{\includegraphics[width=5.in]{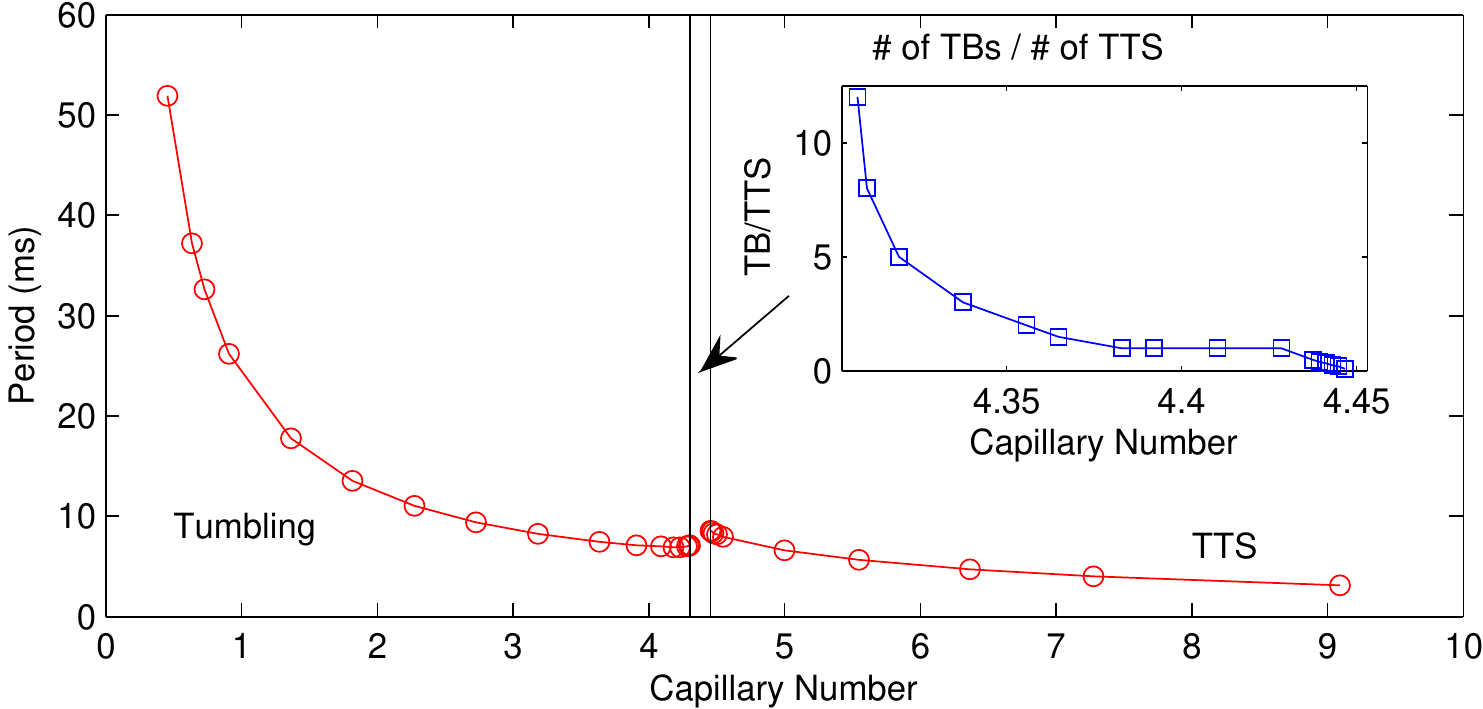}\label{FIG.15.1}}\\
\subfloat[][]{\includegraphics[width=5.in]{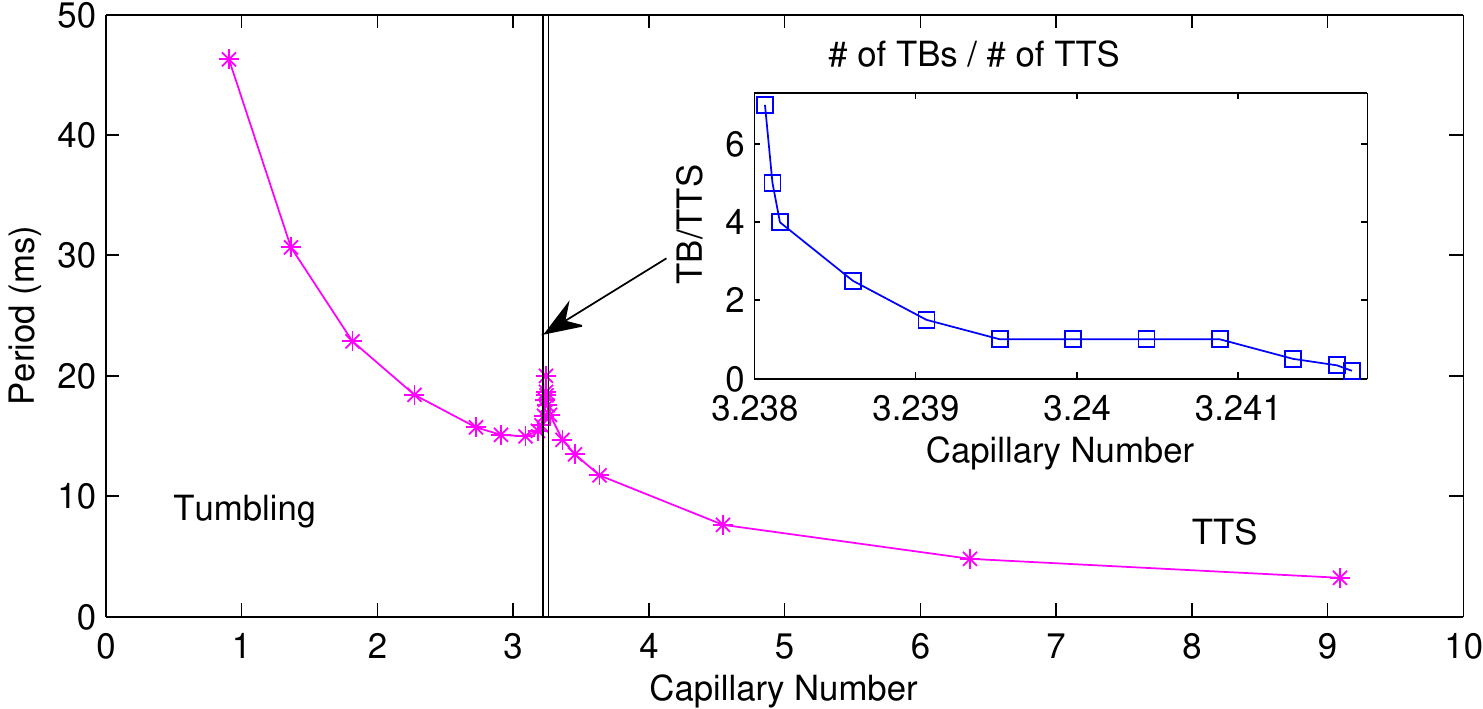}\label{FIG.15.2}}\\
\subfloat[][]{ \includegraphics[width=2.5in]{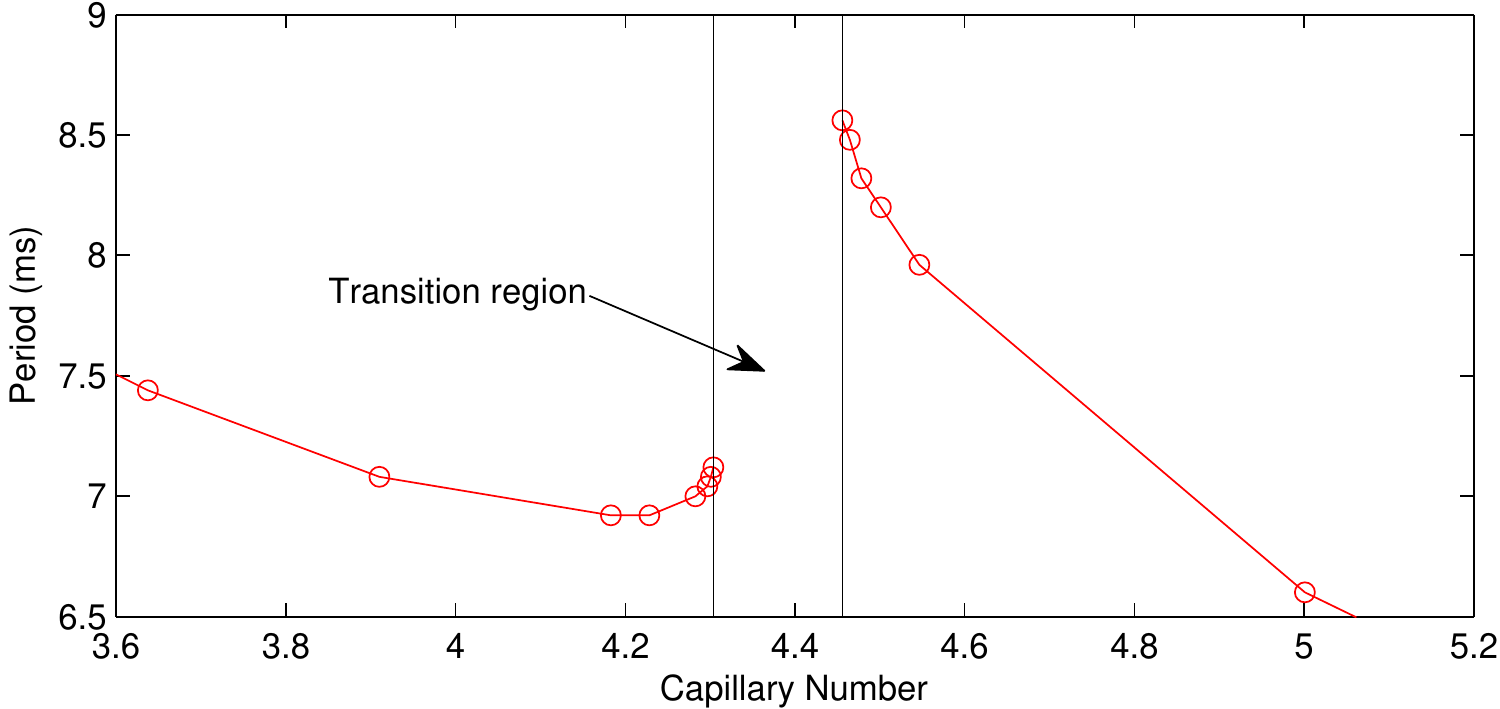}\ 
\includegraphics[width=2.65in]{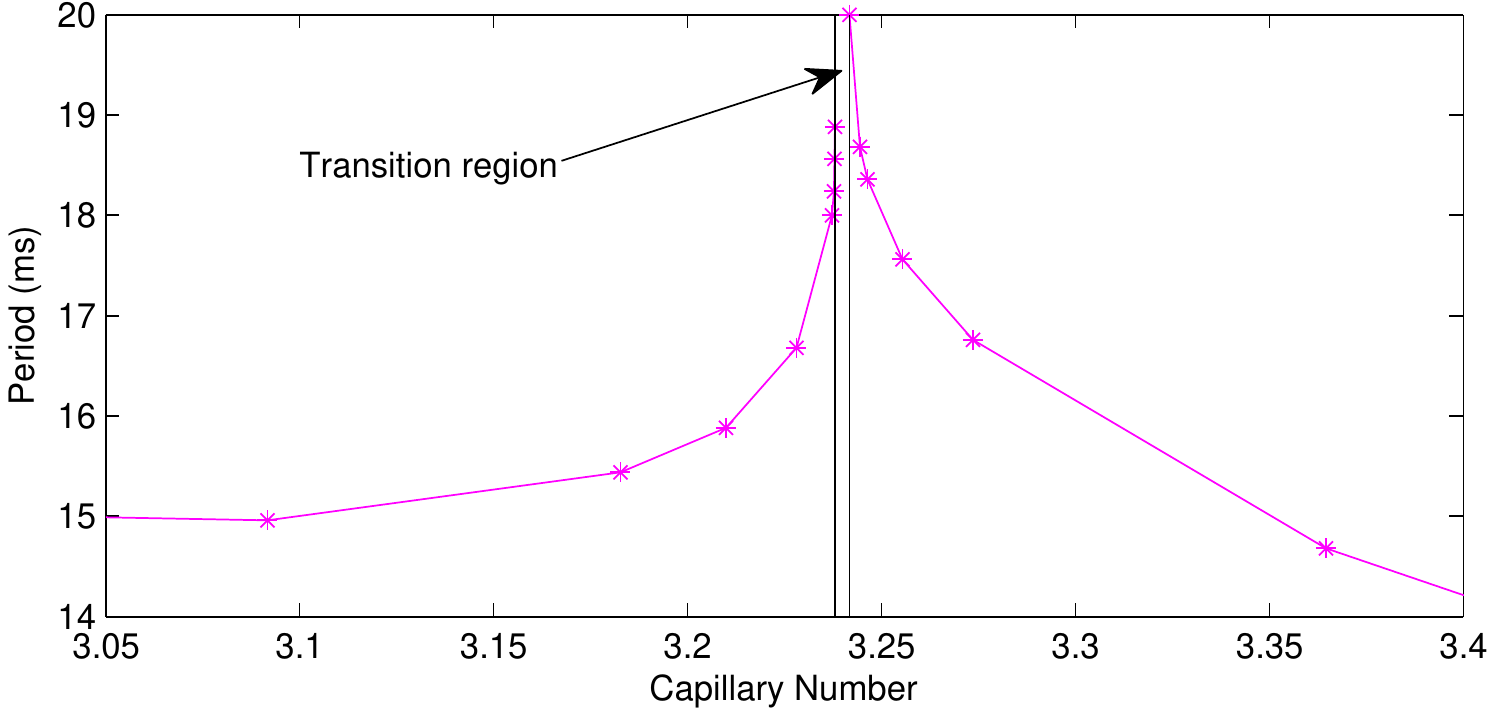} \label{FIG.15.3}}
\end{center}
\caption{(Color online) Histories of period of tumbling and tank-treading with a swinging mode of 
the capsule of $s^*=0.481$ in
\protect\subref{FIG.15.1} a wider channel,
\protect\subref{FIG.15.2} a narrower channel and
\protect\subref{FIG.15.3} the enlargement of the intermittent region
of Fig. \protect\subref{FIG.15.1} (left) and Fig. \protect\subref{FIG.15.2} (right).}\label{FIG.15}
\end{figure}

Here are the details of the intermittent region for the case $\alpha=1$. Since the capsule has a relatively 
strong memory of its shape as $\alpha=1$,  such region is larger than those with weaker non-uniform natural 
state. Actually, according to our simulations, the value of $\alpha$ only effects the critical Capillary 
number and the range of the Capillary number where intermittent behavior occurs.
As shown in Figure \ref{FIG.15.1}, the  capsule performs tumbling motion in a wider channel for
the capillary number $C_a$  less than a critical value of 4.274 and TT with a swinging mode for $C_a$ greater 
than 4.456.  The intermittent region is at  $4.274 < C_a < 4.456$. The embedded sub--figure in 
Figure \ref{FIG.15.1}  corresponds to the contrast ratio between the number of the tumbling and the 
number of the periods of tank--treading with a swinging mode in one cycle of the capsule motion at the 
intermittent region.  When the capillary number is very close to and below the threshold for the transition 
to the pure TT with a swinging mode, the capsule tumbles once after several periods of TT with a swinging mode in one cycle. The number of 
periods of TT with a swinging mode in each cycle decreases with respect to the decreasing of the capillary number, until 
the capsule tumbles once and has one period of TT with a swinging mode alternatively. This should be a relatively stable
dynamics in the intermittent region since we observed it happens over a range of the capillary number. 
Then the capsule performs more tumbling between two consecutive periods of TT with a swinging mode when we keep reducing the 
capillary number, finally when the capillary number is less than the threshold for the intermittent region, 
the capsule just tumbles.   In a narrower channel of dimensions $80\mu$m $\times 10\mu$m, the capsule enters 
the intermittent region at a critical capillary number 3.237, and performs pure TT with a swinging mode for the capillary
number $C_a>3.242$,  which gives a smaller range of the capillary number for the intermittent region 
(see Figure \ref{FIG.15.2}). The embedded sub--figure in Figure \ref{FIG.15.2} also show the capsule behaves
similarly at the intermittent region in a narrower channel. 

\begin{figure}[!pt]
\begin{center}
\includegraphics[width=2.9in]{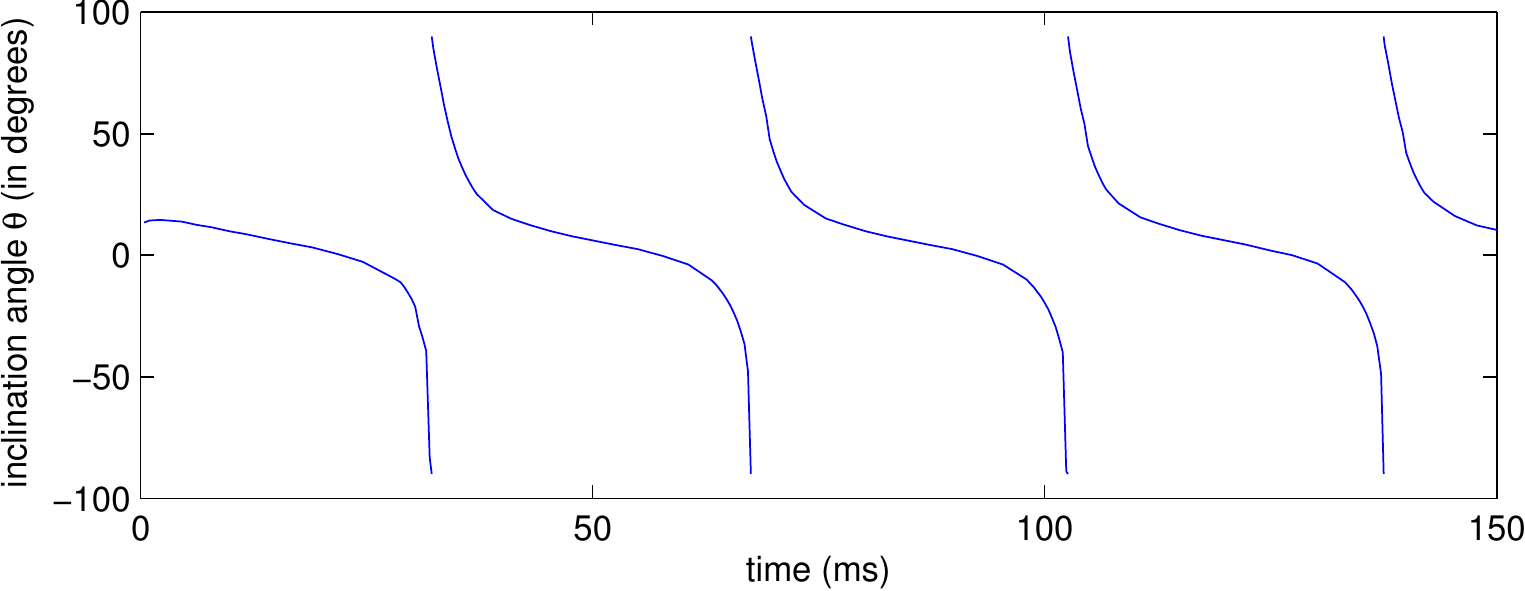} 
\hskip 10pt
\includegraphics[width=2.5in]{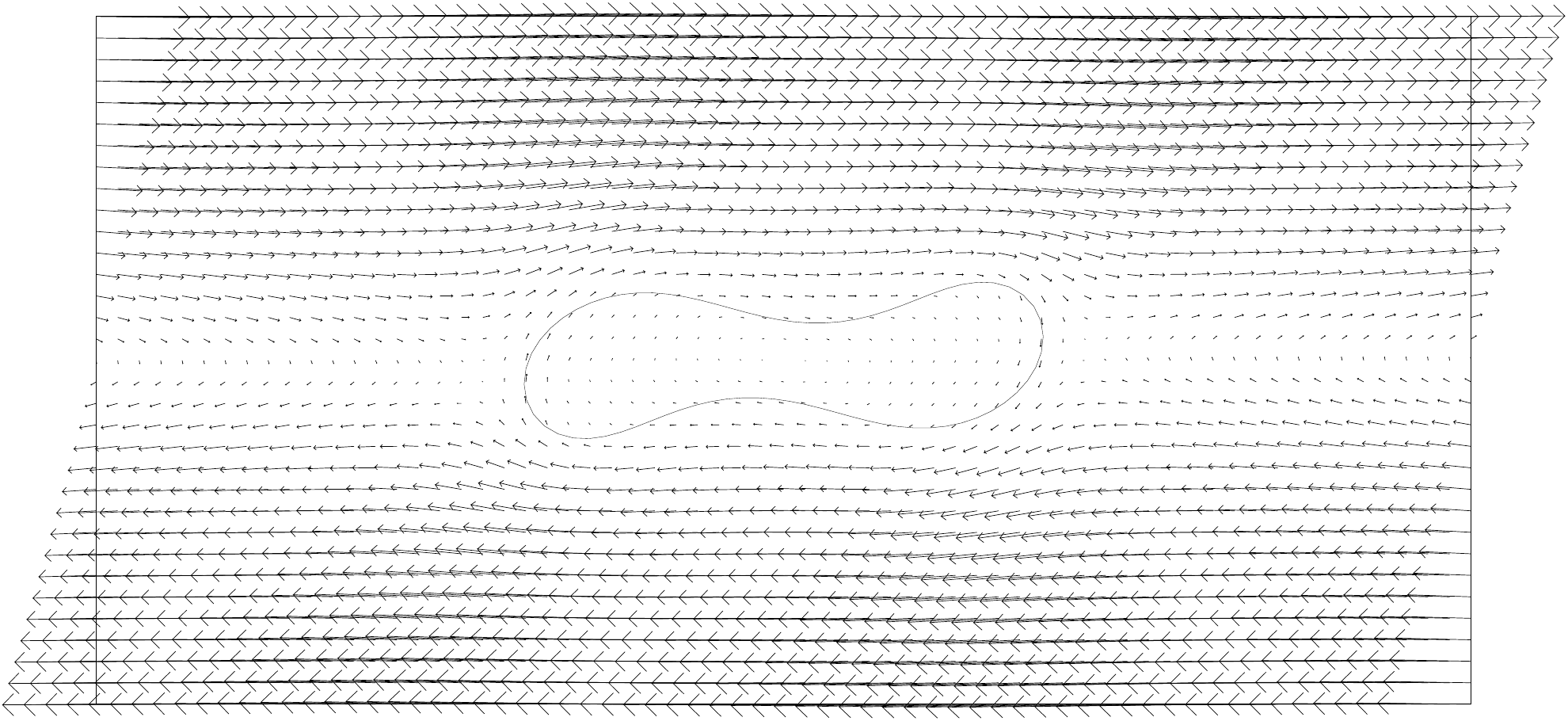}
\end{center}
\caption{(Color online) History of the inclination angle (left) and a velocity filed snapshot (right) of 
the capsule in a wider channel at $Ca$= 4.274.}\label{FIG.16}
\end{figure}

An interesting observation is that the periods of tumbling and TT with a swinging mode have a sharp raise when the 
capsule motion is close to the thresholds as in Figure \ref{FIG.15.3}. For the capillary number 
right above the intermittent region, one of the explanations is that while the capsule changes its 
shape, the membrane particle leaves its natural--state position during the tank-treading motion. 
But the force caused by the bending energy is tending to pull the membrane back to its original 
natural state, which obviously is against the viscous force of flow which would like to push the 
membrane particles moving along the membrane. Hence when the capillary number is just right above the 
intermittent region, the tank--treading motion is slower since the contrast between these two forces 
is not significant when comparing with those of higher capillary number. For the capillary number right 
below the intermittent region, the capsule just 
tumbles and has a shape of long body. In the non-Stokes flow regime, a neutrally buoyant rigid 
particle of elliptic shape in shear flow has a transition from the tumbling motion to the state 
with a fixed inclination angle (i.e., no rotation at all). As the shear rate increases, the 
circulation before and after the long body becomes stronger and then the long body can be  
held by the fluid flow with a fixed inclination angle (see, e.g., \cite{Ding2000,Chen2012}). 
In this paper, the capsule of a long body shape suspended in shear flow is actually a neutrally buoyant entity.
Similarly the capsule slows down its tumbling rotation when the capillary number is less than and closer 
to the threshold for the transition to the intermittent region. Figure \ref{FIG.16} shows the circulation 
of the velocity field before and after the capsule and the history of the inclination angle of the capsule 
in a wider channel  at $Ca$= 4.274 in which the tumbling rotation of the capsule slows down at the 
inclination angle around 5$^\circ$ degrees. Thus the period of tumbling increases as the capillary 
number increases since the strength of the flow field circulation before and after the capsule is increased.

\begin{figure}[ht]
\begin{center}
\subfloat[][]{\includegraphics[width=4.in]{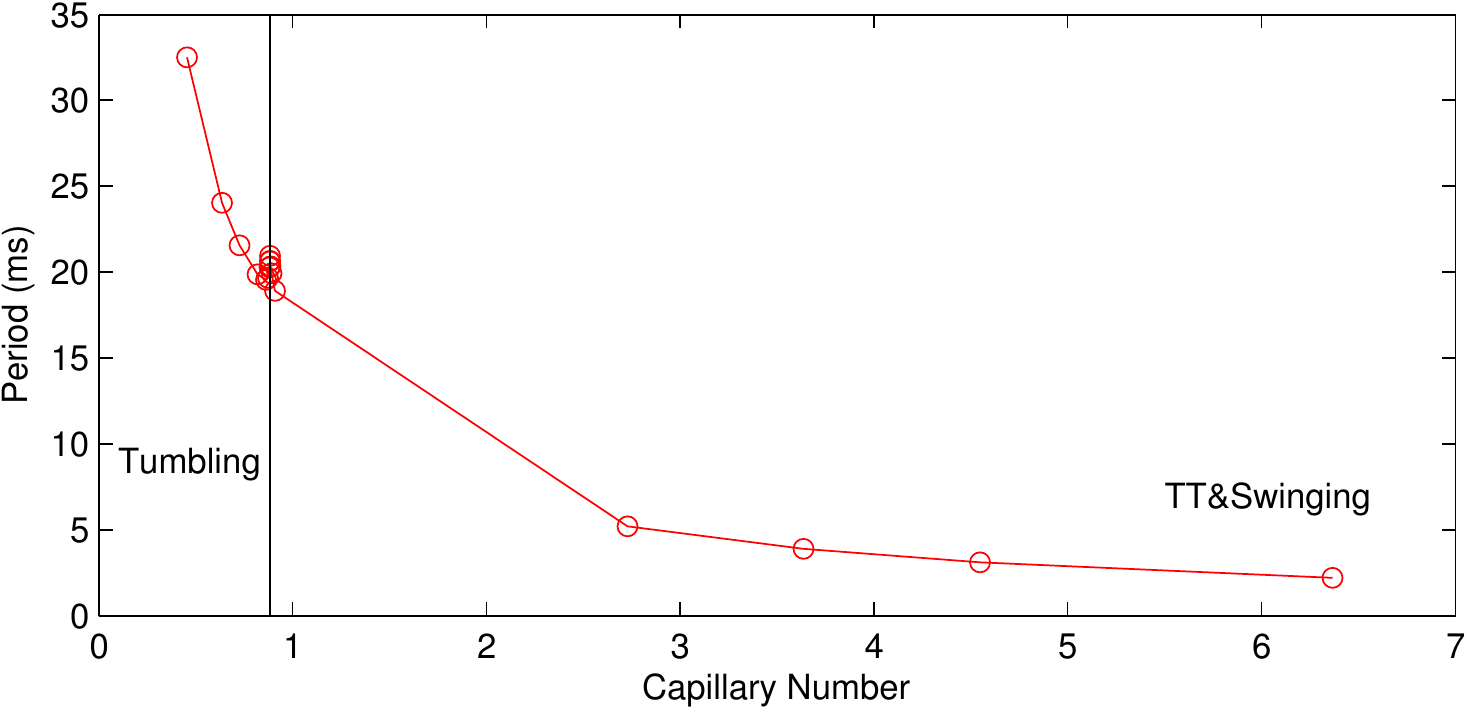}\ \includegraphics[width=1.9in]{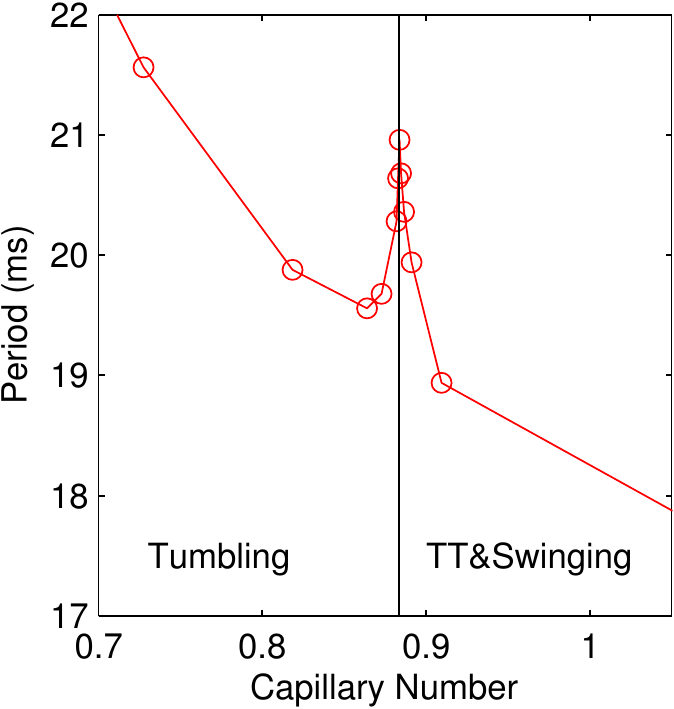}\label{FIG.17.1}}\\
\subfloat[][]{\includegraphics[width=4.in]{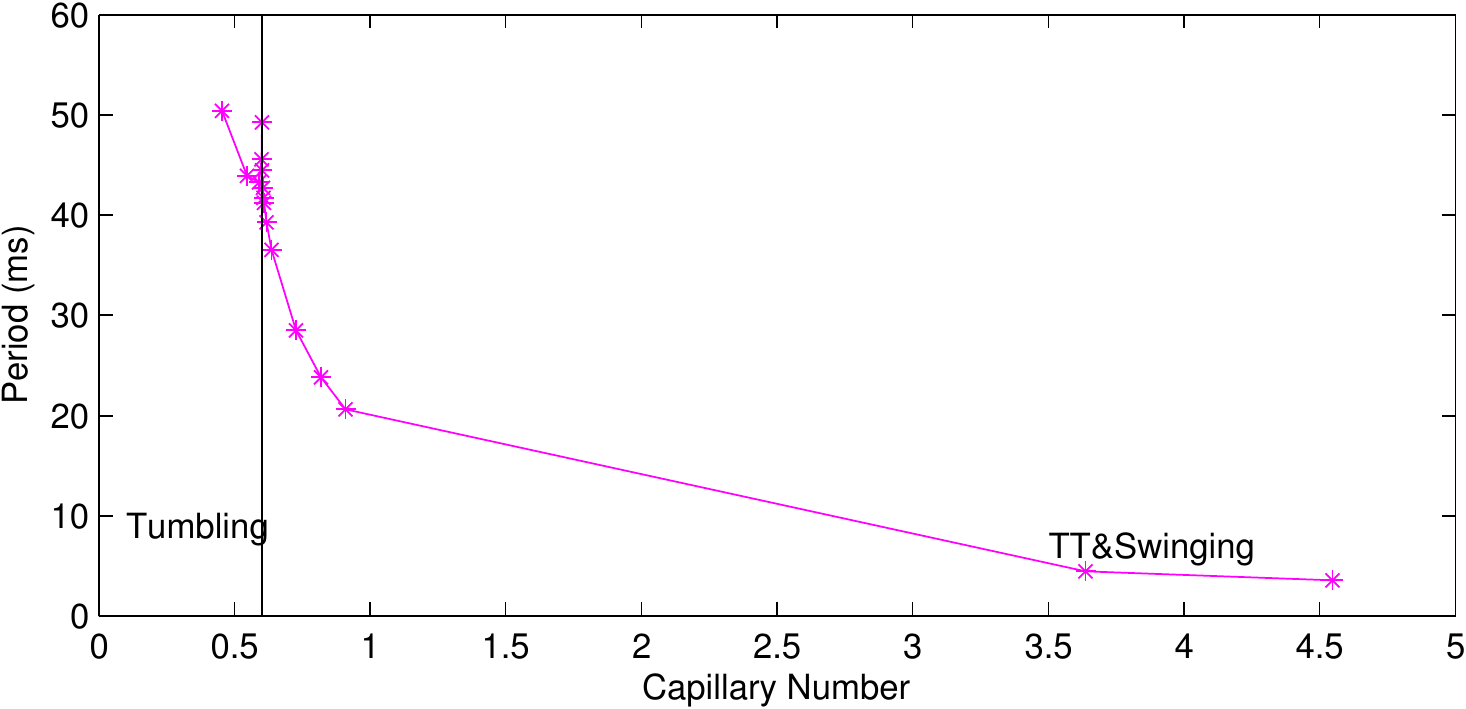}\ \includegraphics[width=1.95in]{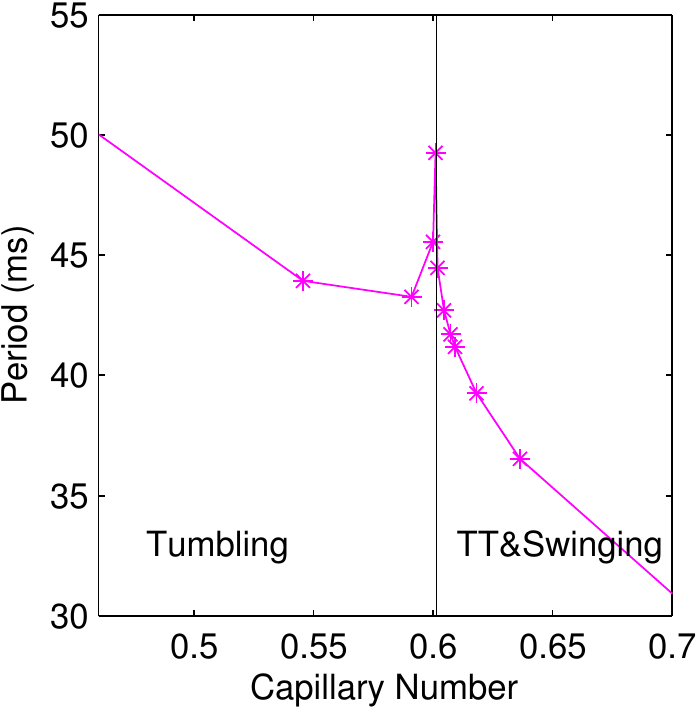} \label{FIG.17.2}}
\end{center}
\caption{(Color online) Histories of period of tumbling and tank--treading of the capsule of $s^*=0.9$ (left) and 
the enlargement of the intermittent region (right) in
\protect\subref{FIG.17.1} a wider channel and
\protect\subref{FIG.17.2} a narrower channel.}\label{FIG.17}
\end{figure}

For the effect of the swelling ratio, we have tested the values of $s^*=0.6$, 0.7, 0.8 and 0.9 and obtained 
that for the capsule of swelling ratio greater than 0.6, it is almost impossible to capture the 
intermittent region computationally since the size of the range of the capillary number for such region
is about zero if it exists. 
In Figure \ref{FIG.17}, results similar to those in Figure \ref{FIG.15} are shown for the case of the capsule of
the swelling ratio $s^*=0.9$, whose shape is about an elliptical shape. The capsule of the swelling ratio 
$s^*$ can be characterized by its excess circumference $\triangle c = 2 \pi (\dfrac{1}{\sqrt{s^*}}-1)$. For the
biconcave shape of $s^*=0.481$, its $\triangle c$ is 2.77638; but the one for an elliptical shape of $s^*=0.9$ is
0.33987. The excess circumference $\triangle c$ is similar to the excess area used in \cite{Young2011}. For the
small values  of  $\triangle c$, we do not expect to obtain the intermittent region.
Our result is consistent with the results obtained by Tsubota {\it et al.} in \cite{Tsubota2010} since the cells used 
in their simulations have the swelling ratio of 0.7.

\subsection{The effect of the nonuniform natural state}\label{sec.5.1}

\begin{figure}[ht]
\begin{center}
\includegraphics[width=4.7in]{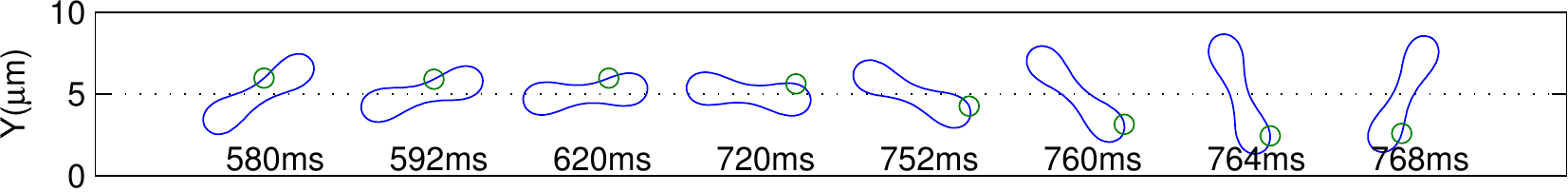}\\
\includegraphics[width=4.7in]{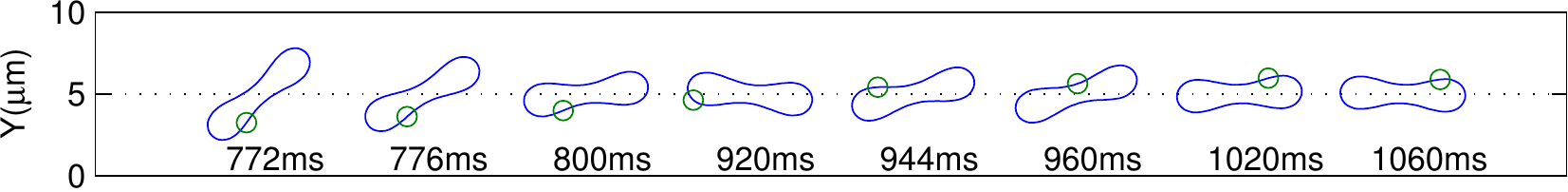}\\
\end{center}
\caption{(Color online) Snapshots  of a tank--treading motion of a capsule with $\alpha=0.05$ 
at the intermittent region in shear flow at $C_a=0.163$.}\label{FIG.18}
\begin{center}
\includegraphics[width=2.7in]{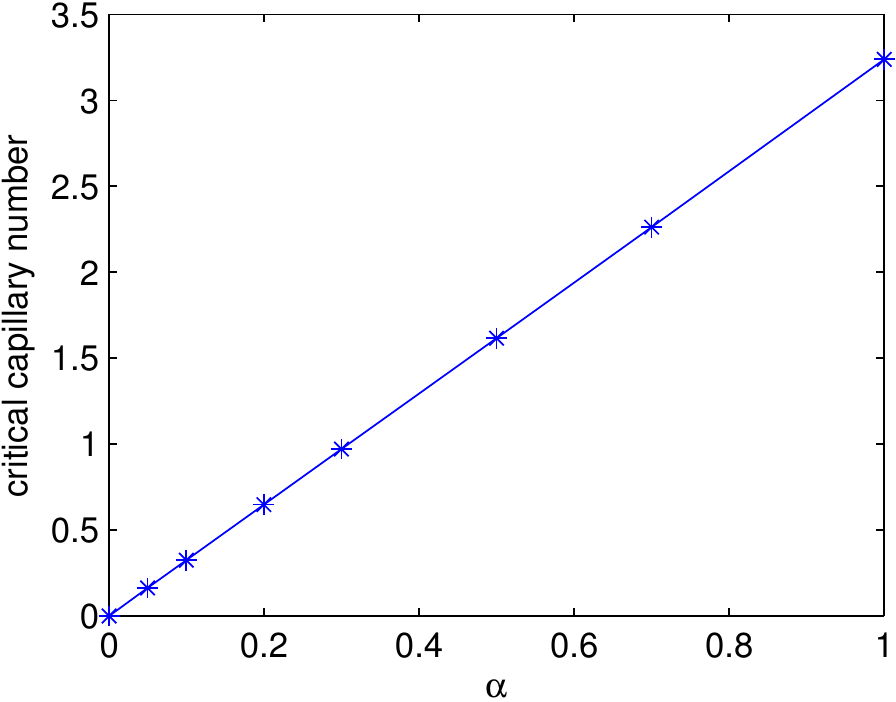} \ 
\includegraphics[width=2.7in]{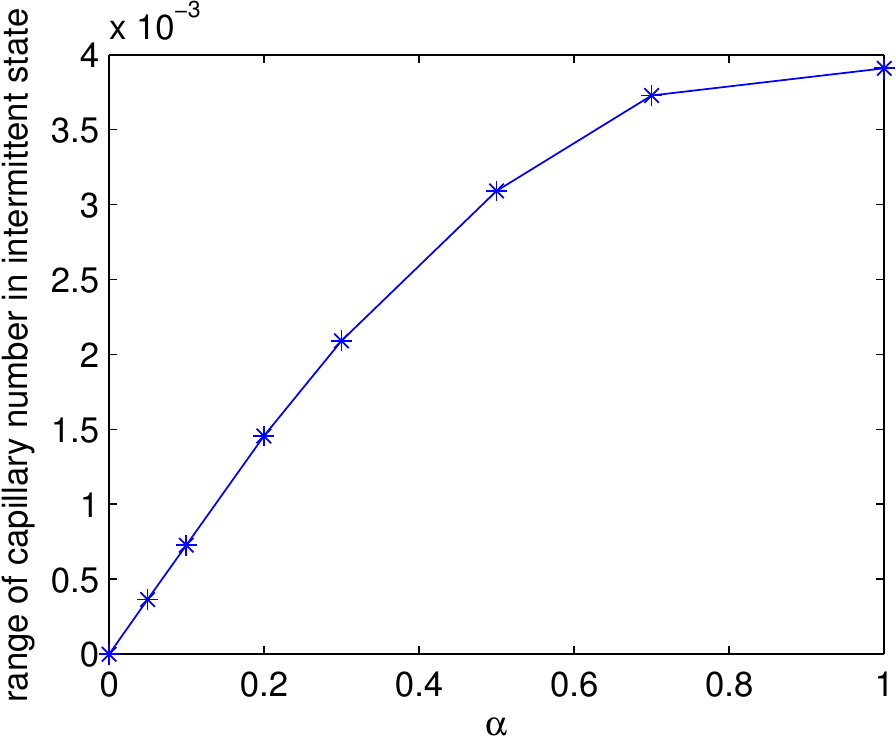}
\end{center}
\caption{(Color online) The  intermittent region   versus the  shape memory coefficient $\alpha$: the plot of 
the critical value of the capillary number for the transition from 
tumbling to the intermittent region   (left) and the size of the  capillary number range for the 
 intermittent region (right).}\label{FIG.19}
\end{figure}

As mentioned in the previous subsection, with lower values of $\alpha$, a capsule does not tend to change
shape as much as capsule with $\alpha=1$. Even in the intermittent region, the capsule prefers keeping the biconcave 
shape (see Figure \ref{FIG.18} for an example), which is unlike the one shown in Figure \ref{FIG.13} for $\alpha=1$.
Figure \ref{FIG.19} shows that the critical value of the capillary number for the
transition from tumbling to the intermittent region   is proportional to the value of $\alpha$, 
this can be explained by defining a weighted bending capillary number with respect to $\alpha$:
$\text{Ca}_{B}^{\alpha}=\mu R_{0}^3\dot{\gamma}/(\alpha k_B l_0)$ which is 
similar to the bending capillary number defined by Tsubota et al. in \cite{Tsubota2010}.
This number represents the relative effect of fluid viscous force versus surface 
tension which comes from  the memory of the reference angles $\{\theta_i^0\}$, and therefore 
 a scale of $\alpha$ should be added. The capsule has tumbling motion when the viscous force from 
outer fluid is less than the force pulling it back to initial position, and TT with a swinging mode when the viscous 
force is larger. Intermittent behavior occurs when  $\text{Ca}_{B}^{\alpha}\sim k$, where $k$ depends on swelling 
ratio and degree of confinement, etc. This implies that $\dot{\gamma}\sim \alpha k_B l_0/(\mu R_{0}^3)$, 
i.e. the capillary number of the intermittency is proportional to the value of $\alpha$ (because the 
capillary number is proportional to the shear rate). The size of   the  capillary number range for 
the intermittent region is increasing as $\alpha$ increases. When $\alpha$ is small, the rate of increasing 
is almost linear. But as $\alpha$ raising up, this rate slows down as the capsule becomes more 
deformable and can change shape accordingly to achieve periodic tumbling or TT with a swinging mode, 
which is more preferred because the energy change is more stable than in intermittent states.

\begin{figure}[tp]
\begin{center}
 \includegraphics[width=4.5in]{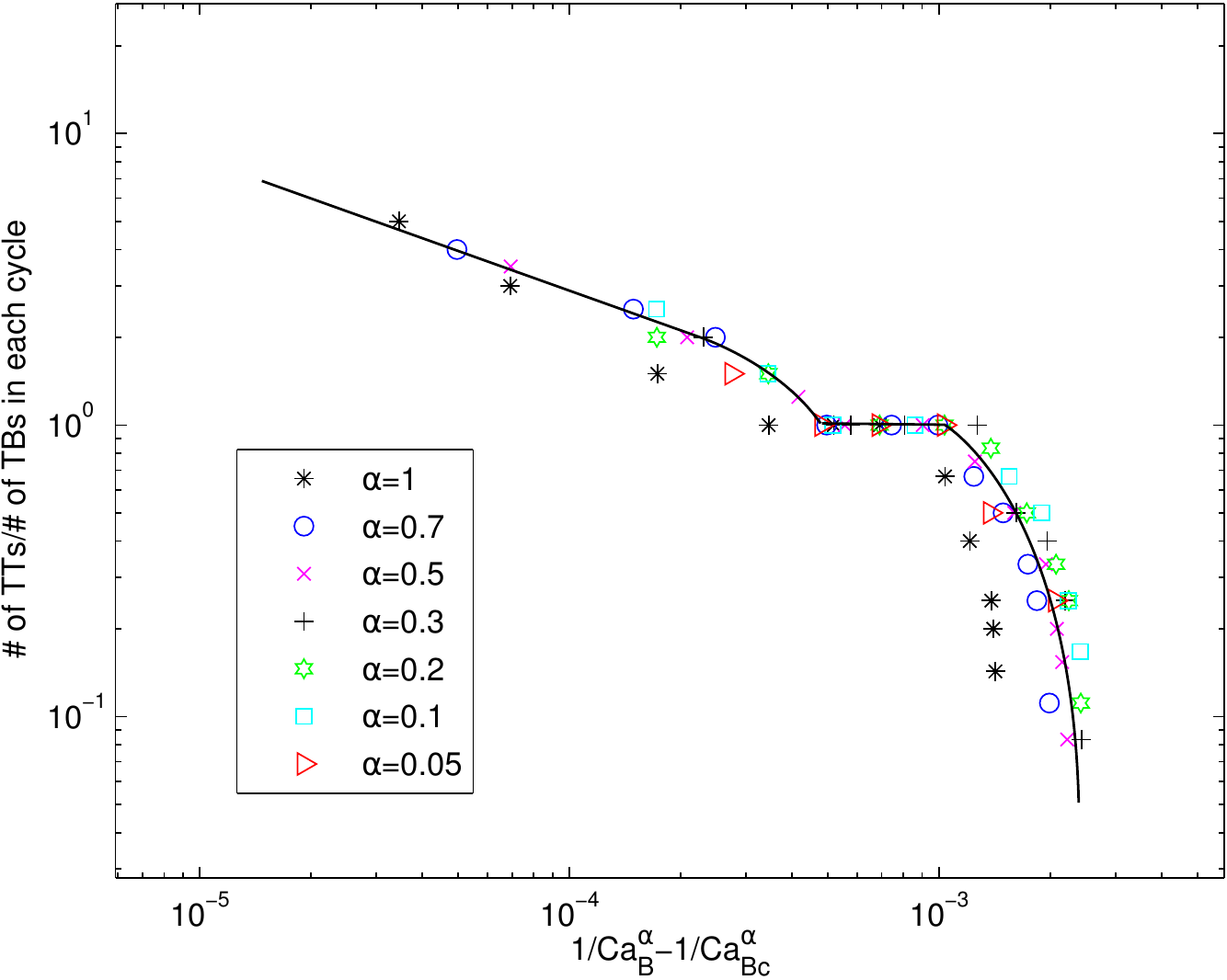} 
 \end{center}
\caption{(Color online) The mixed dynamical behavior of the capsule  with effect of shape memory $\alpha$ varies from 0.05 to 1, in a narrower channel.}\label{FIG.20}
\end{figure}

The mixed dynamical behavior of the capsule in the  intermittent region   with respect to $Ca_B^{\alpha}$ is plotted in 
Figure \ref{FIG.20}, where the y-axis interprets the number of tank--treading with a swinging mode over the number of tumbling 
in each cycle, and x-axis is the difference between $1/Ca_B^{\alpha}$ and its value at the boundary of pure 
tank--treading regime and intermittency.  In the ``more tank--treading than tumbling per cycle'' regime, 
the tendency is almost linear with the slope of $-0.45$ until slope becomes a little sharper when close 
to the ``one tank--treading and one tumbling alternatively'' regime, 
which is consistent with the results obtained by Skotheim and Secomb in \cite{Skotheim2007}.

One major observation through Figure \ref{FIG.20} is that the ``one tumbling and one tank--treading alternatively'' regime
is not eliminated when reducing the effect of nonuniform natural state, even with $\alpha$ values as small as 0.05, where the 
capsule is only little deformable, this suggests this is a relative stable state in the intermittent region.

\section{Conclusions}

In this paper, we have analyzed the dynamics of an inextensible capsule  suspended in two--dimensional 
shear flow under the effect of the nonuniform natural state. Besides the viscosity ratio of the internal 
fluid and external fluid of the capsule, 
the nature state effect also plays a role for having the transition between two well known motions, tumbling  and 
tank-treading   with the long axis oscillating about a fixed inclination angle, when varying 
the shear rate. The intermittent region between tumbling and TT with a swinging mode of the capsule with 
a biconcave rest shape has been obtained in a narrow range of the capillary number. In such region, the 
dynamics of the capsule is a mixture of tumbling and TT with a swinging mode and when having the tumbling motion, 
the membrane tank-tread backward and forward within a small range. As the capillary number 
is very close to and below the threshold for the pure TT with a swinging mode, the capsule tumbles once after 
several TT periods in each cycle. The number of TT periods in one cycle decreases 
with respect to the decreasing of the capillary number, until the capsule has one tumble  and one TT period 
alternatively and such alternating motion exists over a range of the capillary number; and then the capsule
performs more tumbling between two consecutive TT periods when reducing the capillary number further, and finally shows 
pure tumbling. Surprisingly the ``one tumbling and one tank--treading alternatively'' mode is very persistent.

The critical value of the swelling ratio for having the intermittent behavior has been estimated. For those
greater than  0.6 ($\triangle c$=1.82837), it is almost impossible to capture the intermittent behavior 
computationally since the size of the range of the capillary number for such behavior is about 
zero if it exists.  For the small values  of  $\triangle c$ (associated with large swelling
ratio $s^*$), we do not expect to obtain the intermittent region.

An interesting observation is that the period has a sharp raise when the capsule motion is close to the intermittent 
region. For the capillary number right above the intermittent region, the force caused by the bending energy is tending 
to pull the membrane back to its original natural state, which obviously is against the viscous force of flow 
which would like to push the membrane particles moving along the membrane. Hence when the capillary number is right 
above the intermittent region, the tank--treading motion is slower since the contrast between these two forces is not 
significant. For the capillary number right below the intermittent region, the capsule is a neutrally buoyant entity
and slows down its tumbling rotation when the capillary number is less than and closer to the threshold for the 
intermittent region, which is closely related to the case of a neutrally buoyant rigid particle in shear flow. 

Another observation is that in the intermittent region, the change of contrast between numbers of 
tumbling and tank--treading is not linear, especially when the weighted bending capillary number is close to or in
the range of a relatively stable ``one tumbling and one tank--treading per cycle'' mode, we discussed this 
phenomenon may result from the deformability of capsule, even with the value of $\alpha$ as small as 0.05,
the capsule may still have some invisible shape changes and is able to perform one tumbling one tank--treading 
in each cycle within a range of weighted bending capillary number.

\section*{Acknowledgments}
This work is supported by an NSF grant DMS--0914788. We acknowledge the helpful comments of James Feng,
Ming-Chih Lai and Sheldon X. Wang.

\end{document}